\documentclass[aps,prd,showpacs,preprintnumbers,nofootinbib]{revtex4}
\usepackage{graphicx}
\usepackage{amsmath,amssymb}
\usepackage{mathrsfs}
\usepackage{bm}

\preprint{CHIBA-EP-240, 2019.07.08}

\begin{document}
\title{
Dyon in the $SU(2)$ Yang--Mills theory 
with a gauge-invariant gluon mass toward quark confinement
}

\author{Shogo Nishino}
\email{shogo.nishino@chiba-u.jp}

\author{Kei-Ichi Kondo}

\affiliation{Department of Physics,  
Graduate School of Science, 
Chiba University, Chiba 263-8522, Japan
}

\begin{abstract}
In the previous paper, we have shown the existence of magnetic monopoles in the pure $SU(2)$ Yang--Mills theory with a gauge-invariant mass term for the gluon field being introduced.
In this paper, we extend our previous construction of magnetic monopoles to obtain dyons with both magnetic and electric charges.
In fact, we solve under the static and spherically symmetric ansatz the field equations of the $SU(2)$ ``complementary'' gauge-scalar model, which is the $SU(2)$ Yang--Mills theory coupled to a single adjoint scalar field whose radial degree of freedom is eliminated. 
We show that the novel dyon solution can be identified with the gauge field configuration of a dyon with a minimum magnetic charge in the massive Yang--Mills theory. 
Moreover, we compare the dyon of the massive Yang--Mills theory obtained in this way with the Julia--Zee dyon in the Georgi--Glashow gauge-Higgs  scalar model  and the dyonic extension of the Wu--Yang magnetic monopole in the pure Yang--Mills theory. 
Finally, we identify the novel dyon solution found in this paper with a dyon configuration on $S^1 \times \mathbb{R}^3$ space with nontrivial holonomy and propose to use it to understand the confinement/deconfinement phase transition in the Yang--Mills theory at finite temperature, instead of using the dyons constituting the Kraan--van Baal--Lee--Lu caloron.

\end{abstract}

\maketitle

\section{Introduction}
Quark confinement is a long-standing problem to be solved in the framework of quantum chromo-dynamics (QCD). 
One of the most promising scenarios for quark confinement is the dual superconductivity picture \cite{dual-superconductor} for QCD vacuum.
For this hypothesis to be realized, the existence of the relevant magnetic objects and their condensations are indispensable.  
In the gauge-scalar model, it is indeed well known that the 't Hooft--Polyakov magnetic monopole \cite{tHP} exists as a  topological soliton solution of the field equations of the Georgi--Glashow gauge-Higgs scalar model in which  a single scalar field belongs to the adjoint representation of the gauge group $SU(2)$. 
However, such a scalar field is not   introduced in the original action of QCD. 
Thus, we are faced with the problem of showing the existence of magnetic monopoles in the Yang--Mills theory in absence of the scalar field.
See e.g., \cite{KKSS} for a review. 

In a previous paper \cite{Nishino}, nevertheless, we have succeeded to obtain the magnetic monopole {\it configuration} with a nontrivial magnetic charge in the pure $SU(2)$ Yang--Mills theory when a gauge-invariant mass term for the gauge field \cite{Kondo2016} is introduced without any scalar field.  
We call this theory the {\it massive Yang--Mills theory} and call the resulting magnetic monopole the \textit{Yang--Mills magnetic monopole}.
This result follows from the recent proposal for obtaining the gauge field configurations in the pure Yang--Mills theory from solutions of field equations in the ``complementary'' gauge-scalar model \cite{Kondo2016} in which the radial degree of freedom of a single adjoint scalar field is frozen. 
The gauge-invariant mass term is obtained through change of variables and a gauge-independent description \cite{Kondo2016,Kondo2018} of the Brout--Englert--Higgs (BEH) mechanism \cite{BEH}, which neither relies on the spontaneous breaking of gauge symmetry nor on the assumption of a nonvanishing vacuum expectation value of the scalar field.

As first shown by Julia and Zee \cite{Julia--Zee}, 
it is possible to provide magnetic monopoles with electric charges, in fact, a 't Hooft--Polyakov magnetic monopole \cite{tHP} can have the electric charge in addition to the magnetic charge, which is called a {\it dyon} \cite{textbooks}.
We show in this paper that the dyon configuration exists in the pure $SU(2)$ Yang--Mills theory with a gauge-invariant mass term of the gauge field, i.e., the massive Yang--Mills theory.
We call this dyon the {\it Yang--Mills dyon}.
In fact, we solve under the static and spherically symmetric ansatz the field equations of the $SU(2)$ ``complementary'' gauge-scalar model 
and obtain a gauge field configuration for a dyon with a minimum magnetic charge in the massive  Yang--Mills theory. 
In particular, we compare the dyon obtained in this way in the massive Yang--Mills theory with the Julia--Zee dyon in the Georgi--Glashow gauge-Higgs model and the dyonic extension of the Wu--Yang magnetic monopole \cite{Wu--Yang} in the pure Yang--Mills theory.

It is well known that topological solitons play very important roles in non-perturbative investigations of the Yang--Mills theory.
For example, the Kraan--van Baal--Lee--Lu (KvBLL) caloron (instanton) \cite{KvBLL} has been extensively used to reproduce the confinement/deconfinement phase transition in the Yang--Mills theory at finite temperature \cite{Diakonov}.
The KvBLL caloron is a topological soliton solution of the (anti-)self-dual equation of the $SU(2)$ Yang--Mills theory on $S^{1} \times \mathbb{R}^{3}$ space with a nontrivial instanton charge as the topological invariant, which consists of Bogomol'nyi--Prasad--Sommerfield (BPS) dyons having both electric and magnetic charges with a nontrivial holonomy at spatial infinity.
In contrast, our dyon solution is a non-BPS solution of the (non-self-dual) field equation of the ``complementary'' gauge-scalar model.
Our dyon has the nonvanishing asymptotic value as the nontrivial holonomy at spatial infinity, which is a common property to be comparable with (anti-)self-dual dyons as the constituents of the KvBLL calorons.
The dyon solution of the ``complementary'' gauge-scalar model is identified with the dyon {\it configuration} of the massive Yang--Mills theory with a gauge-invariant mass term of the gauge field without any scalar field, which is regarded as the low-energy effective model of the Yang--Mills theory with a mass gap.
Thus, we can propose another scenario for investigating the confinement/deconfinement phase transition in the Yang--Mills theory at finite temperature based on the novel non-self-dual dyon solution found in this paper.

This paper is organized as follows.
In section I\hspace{-.1em}I, we review the procedure \cite{Kondo2016} for obtaining the massive $SU(2)$ Yang--Mills theory from the ``complementary'' $SU(2)$ gauge-adjoint scalar model.
In section I\hspace{-.1em}I\hspace{-.1em}I, we give a brief review of the Julia--Zee dyon in the ($3+1$)-dimensional Minkowski spacetime. 
In section I\hspace{-.1em}V, we give a Yang--Mills dyon solution of an $SU(2)$  gauge-scalar model in which the radial degree of freedom of a single adjoint scalar field is frozen. 
In section V, we give the decomposition of the gauge  field for the Yang--Mills dyon solution. 
In section V\hspace{-.1em}I, we give the magnetic and electric fields of the Yang--Mills dyon solution.
In section V\hspace{-.1em}I\hspace{-.1em}I, we give the energy density and the static mass of a Yang--Mills dyon. 
In section V\hspace{-.1em}I\hspace{-.1em}I\hspace{-.1em}I, we discuss how the Yang--Mills dyon can be responsible for confinement/deconfinement phase transition of the Yang--Mills theory at finite temperature. 
The final section I\hspace{-.1em}X is devoted to conclusion and discussion.

\section{The massive Yang--Mills theory ``complementary'' to the gauge-adjoint scalar model}

In this section, we review the procedure \cite{Kondo2016} for obtaining the massive $SU(2)$ Yang--Mills theory from the ``complementary'' $SU(2)$ gauge-adjoint scalar model. 
 For this purpose, we introduce the two products for the Lie-algebra valued fields $\mathcal{P} := \mathcal{P}^{A} T_{A}$ and $\mathcal{Q} = \mathcal{Q}^{A} T_{A}$ ($A=1,2,3$):
\begin{align}
\mathcal{P} \cdot \mathcal{Q} := & \mathcal{P}^{A} \mathcal{Q}^{A} , \\
\mathcal{P} \times \mathcal{Q} := & \epsilon^{A B C} T_{A} \mathcal{P}^{B} \mathcal{Q}^{C}
,
\end{align}
where $T_{A}$ are the generators of the Lie algebra $\mathfrak{su} (2)$ of the group $SU(2)$.
We choose the Hermitian basis $T_{A}$    by using the Pauli matrices $\sigma_{A}$ $(A=1,2,3)$ as
\begin{equation}
T_{A} = \frac{1}{2} \sigma_{A}
.
\end{equation}

We introduce the Hermitian $SU(2)$ gauge field $\mathscr{A}_{\mu}(x)$ and the Hermitian  scalar field $\bm{\phi} (x)$ by 
$\mathscr{A}_{\mu} (x) := \mathscr{A}_{\mu}^{A} (x)  {T}_{A}$ 
and 
$\phi (x) = \phi^{A} (x)  {T}_{A}$
as the Lie algebra $\mathfrak{su} (2)$ valued fields.  
Then, we introduce the $SU(2)$ gauge-adjoint scalar model by the Lagrangian density
\begin{equation}
\mathscr{L}_{\rm YM} = - \frac{1}{4} \mathscr{F}_{\mu \nu} \cdot \mathscr{F}^{\mu \nu} + \frac{1}{2} \left( \mathscr{D}_{\mu} [\mathscr{A}] \bm{\phi} \right) \cdot \left( \mathscr{D}^{\mu} [\mathscr{A}] \bm{\phi} \right) + u \left( \bm{\phi} \cdot \bm{\phi} - v^{2} \right)
,
\label{L_YM}
\end{equation}
where $u$ is the Lagrange multiplier field to incorporate the radially fixing constraint
\begin{equation}
\bm{\phi} (x) \cdot \bm{\phi} (x) = v^{2}
.
\label{constraint}
\end{equation}
Here $\mathscr{F}_{\mu \nu}$ denotes the field strength of the $SU(2)$ gauge field $\mathscr{A}_{\mu}$ and $\mathscr{D}_{\mu} [\mathscr{A}] \bm{\phi}$  is the covariant derivative of the scalar field $\bm{\phi} (x)$ defined by
\begin{align}
\mathscr{F}_{\mu \nu} (x) := & \partial_{\mu} \mathscr{A}_{\nu} (x) - \partial_{\nu} \mathscr{A}_{\mu} (x) - g \mathscr{A}_{\mu} (x) \times \mathscr{A}_{\nu} (x) , \\
\mathscr{D}_{\mu} [\mathscr{A}] \bm{\phi} (x) := & \partial_{\mu} \bm{\phi} (x) - g \mathscr{A}_{\mu} (x) \times \bm{\phi} (x) 
.
\end{align}

First of all, we construct a composite vector boson field $\mathscr{X}_{\mu} (x)$ from $\mathscr{A}_{\mu} (x)$ and $\hat{\bm{\phi}} (x)$ as
\begin{equation}
  \mathscr{X}_{\mu} (x) := g^{-1} \hat{\bm{\phi}} (x) \times \mathscr{D}_{\mu} [\mathscr{A}] \hat{\bm{\phi}} (x)
,
\label{decomposition}
\end{equation}
by introducing the normalized scalar field
\begin{equation}
\hat{\bm{\phi}} (x) := \frac{1}{v} \bm{\phi} (x) 
,
\end{equation}
which can be identified with the color direction field $\bm{n} (x)$ in the gauge-covariant field decomposition of the gauge field, see  \cite{KKSS, Kondo-Murakami-Shinohara}.
Notice that $\mathscr{X}_{\mu} (x)$ transforms according to the adjoint representation under the gauge transformation $U (x) \in SU(2)$:
\begin{equation}
\mathscr{X}_{\mu} (x) \to \mathscr{X}^{\prime}_{\mu} (x) = U (x) \mathscr{X}_{\mu} (x) U^{\dagger} (x)
.
\label{gauge_transf_X}
\end{equation}
Then the kinetic term of the scalar field is identical to the mass term of the vector field $\mathscr{X}_{\mu} (x)$:
\begin{equation}
\frac{1}{2} \mathscr{D}^{\mu} [\mathscr{A}] \bm{\phi} \cdot \mathscr{D}_{\mu} [\mathscr{A}] \bm{\phi} (x) = \frac{1}{2} M_{\mathscr{X}}^{2} \mathscr{X}^{\mu} (x) \cdot \mathscr{X}_{\mu} (x), \ \ \ 
M_{\mathscr{X}} := g v
,
\end{equation}
as long as the radial degree of freedom of the scalar field is fixed \cite{Kondo2016}.
It is obvious that the obtained mass term of $\mathscr{X}_{\mu} (x)$ is gauge-invariant by observing (\ref{gauge_transf_X}).
Therefore, $\mathscr{X}_{\mu} (x)$ can be identified with the  massive component without breaking the original gauge symmetry.
This gives a gauge-independent definition of the massive modes of the gauge field in the operator level.
It should be emphasized that we do not need to choose a specific vacuum of $\bm{\phi} (x)$ and hence no spontaneous symmetry breaking of the gauge symmetry occurs.

By using the definition of the massive vector field $\mathscr{X}_{\mu} (x)$, the original gauge field $\mathscr{A}_{\mu} (x)$ is separated into two pieces \cite{Kondo2016,KKSS}:
\begin{equation}
\mathscr{A}_{\mu} (x) := \mathscr{V}_{\mu} (x) + \mathscr{X}_{\mu} (x)
,
\end{equation}
where the field $\mathscr{V}_{\mu} (x)$ can be written in terms of $\mathscr{A}_{\mu} (x)$ and $\hat{\bm{\phi}} (x)$:
\begin{equation}
 \mathscr{V}_{\mu} (x) =  c_{\mu} (x) \hat{\bm{\phi}} (x) + g^{-1} \partial_{\mu} \hat{\bm{\phi}} (x) \times \hat{\bm{\phi}} (x) , \ \ \ 
c_{\mu} (x) = \mathscr{A}_{\mu} (x) \cdot \hat{\bm{\phi}} (x)
.
\end{equation}
Here $\mathscr{V}_{\mu} (x)$ is called the restricted (or residual) part which is expected to give the dominant contribution to quark confinement, while $\mathscr{X}_{\mu} (x)$ is called the remaining (or broken) part which is identified with the massive mode which is expected to decouple in the low-energy or long-distance region.

Then, we regard a set of field variables $\{ c_{\mu} (x) , \mathscr{X}_{\mu} (x) , \hat{\bm{\phi}} (x) \}$ as obtained from $\{ \mathscr{A}_{\mu} (x) , \hat{\bm{\phi}} (x) \}$ based on a change of variables:
\begin{equation}
\{ \mathscr{A}_{\mu} (x) , \hat{\bm{\phi}} (x) \} \to
\{ c_{\mu} (x) , \mathscr{X}_{\mu} (x) , \hat{\bm{\phi}} (x) \} 
,
\end{equation}
and identify $c_{\mu} (x) , \mathscr{X}_{\mu} (x)$ and $\hat{\bm{\phi}} (x)$ with the fundamental field variables for describing the massive Yang--Mills theory anew, which means that we should perform the quantization with respect to the variables $\{ c_{\mu} (x) , \mathscr{X}_{\mu} (x) , \hat{\bm{\phi}} (x) \}$ appearing in the path-integral measure.

In the gauge-scalar model, $\mathscr{A}_{\mu} (x)$ and $\hat{\bm{\phi}} (x)$ are independent field variables. However, the Yang--Mills theory should be described by $\mathscr{A}_{\mu} (x)$ alone.
Hence the scalar field $\bm{\phi} (x)$ must be supplied by the gauge field $\mathscr{A}_{\mu} (x)$ due to the strong interactions, or in other words, $\bm{\phi} (x)$ should be given as a functional of the gauge field $\mathscr{A}_{\mu} (x)$.

Moreover, the independent degrees of freedom of the original gauge field $\mathscr{A}_{\mu}^{A} (x)$ in the pure $SU (2)$ Yang--Mills theory in $D$-dimensional space-time are $[ \mathscr{A}_{\mu}^{A} (x) ] = 3 \times D = 3D$.
Here, we have omitted the infinite degrees of freedom of the space-time points.
On the other hand, the new field variables have independent degrees of freedom: $[ c_{\mu} (x) ] = D$, $[ \hat{\bm{\phi}} (x) ] = 2$, $[ \mathscr{X}_{\mu}^{A} (x) ] = 2 \times D = 2D$, 
where the massive vector field $\mathscr{X}_{\mu} (x)$ obeys the condition:
\begin{equation*}
\mathscr{X}_{\mu} (x) \cdot \hat{\bm{\phi}} (x) = 0
.
\end{equation*}

We can therefore observe that the theory with the new field variables has two extra degrees of freedom if we wish to obtain the (pure) Yang--Mills theory from the ``complementary'' gauge-scalar model. 
These extra degrees of freedom are eliminated by imposing the two constraints which we call the {\it reduction condition}.
We choose e.g., the reduction condition 
\begin{equation}
\bm{\chi} (x) := \hat{\bm{\phi}} (x) \times \mathscr{D}^{\mu} [\mathscr{A}] \mathscr{D}_{\mu} [\mathscr{A}] \hat{\bm{\phi}} (x) = 0
.
\label{reduction2}
\end{equation}
The reduction condition indeed eliminates the two extra degrees of freedom introduced by the radially fixed scalar field into the Yang--Mills theory, since
\begin{equation}
\bm{\chi} (x) \cdot \hat{\bm{\phi}} (x) = 0
.
\end{equation}

Following the Faddeev--Popov procedure, we insert the unity to the functional integral to incorporate the reduction condition:
\begin{equation}
1 = \int \mathcal{D} \bm{\chi}^{\theta} \ \delta \left( \bm{\chi}^{\theta} \right) 
= \int \mathcal{D} \bm{\theta} \ \delta \left( \bm{\chi}^{\theta} \right) \Delta^{\rm red}
,
\end{equation}
where $\bm{\chi}^{\theta} := \bm{\chi} [ \mathscr{A} , \bm{\phi}^{\theta} ]$ is the reduction condition written in terms of $\mathscr{A}_{\mu} (x)$ and $\bm{\phi}^{\theta}$ which is the local rotation of $\bm{\phi} (x)$ by $\bm{\theta} = \bm{\theta} (x) = \theta^{A} (x) T_{A}$ and $\Delta^{\rm red} := \det \left( \frac{\delta \bm{\chi}^{\theta}}{\delta \bm{\theta}} \right)$ denotes the Faddeev--Popov determinant associated with the reduction condition $\bm{\chi} = 0$.
Then, we write the vacuum-to-vacuum amplitude of the gague-scalar model subject to the reduction condition is translated into the massive Yang--Mills theory with the gauge-invariant mass term of the field $\mathscr{X}$ as  
\begin{align}
Z = & \int \mathcal{D} \hat{\bm{\phi}} \mathcal{D} \mathscr{A} \ \delta \left( \bm{\chi} \right) \Delta^{\rm red} \exp \left\{ i S_{\rm YM} [\mathscr{A}] + i S_{\rm kin} [ \mathscr{A} , \bm{\phi}] \right\} \nonumber\\
= & \int \mathcal{D} \hat{\bm{\phi}} \mathcal{D} c \mathcal{D} \mathscr{X} \ J \delta \left( \widetilde{\bm{\chi}} \right) \widetilde{\Delta}^{\rm red} \exp \left\{ i S_{\rm YM} [ \mathscr{V} + \mathscr{X}] + i S_{m} [ \mathscr{X} ] \right\}
,
\label{path_integral}
\end{align}
where the Jacobian $J$ associated with the change of variables is equal to one, $J = 1$ \cite{KKSS}.
Therefore, we obtain the massive Yang--Mills theory which keeps the original gauge symmetry:
\begin{equation}
\mathscr{L}_{\rm mYM} = - \frac{1}{4} \mathscr{F}_{\mu \nu} [ \mathscr{V} + \mathscr{X} ] \cdot \mathscr{F}^{\mu \nu} [ \mathscr{V} + \mathscr{X} ] + \frac{1}{2} M_{\mathscr{X}}^{2} \mathscr{X}_{\mu} \cdot \mathscr{X}^{\mu}
, \ \ \ M_{\mathscr{X}} : = g v > 0 .
\label{mYM}
\end{equation}
The obtained massive Yang--Mills theory indeed has the same degrees of freedom as the usual Yang--Mills theory because the massive vector boson $\mathscr{X}_{\mu} (x)$ is constructed by combining the original gauge field $\mathscr{A}_{\mu} (x)$ and the normalized scalar field $\hat{\bm{\phi}}(x)$ where $\hat{\bm{\phi}} (x)$ is now a (complicated) functional of $\mathscr{A}_{\mu} (x)$ obtained by solving the reduction condition (\ref{reduction2}).

It should be remarked that the solutions of the field equations of the gauge-scalar model satisfy the reduction condition automatically, although the converse is not true \cite{Kondo2016}. 
The field equations besides the constraint equation (\ref{constraint}) are obtained as
\begin{align}
\mathscr{D}^{\mu} [\mathscr{A}] \mathscr{F}_{\mu \nu} - g  \bm{\phi} \times \mathscr{D}_{\nu} [\mathscr{A}] \bm{\phi}  = & 0 , 
\label{YM_eq_gauge}
\\
\mathscr{D}^{\mu} [\mathscr{A}] \mathscr{D}_{\mu} [\mathscr{A}] \bm{\phi}  - 2 u \bm{\phi} = & 0
.
\label{YM_eq_scalar}
\end{align}
We take the inner product of (\ref{YM_eq_scalar}) and $\bm{\phi} (x)$ and use (\ref{constraint}) to obtain
\begin{equation}
u = \frac{1}{2 v^{2}} \bm{\phi} \cdot \left( \mathscr{D}^{\mu} [\mathscr{A}] \mathscr{D}_{\mu} [\mathscr{A}] \bm{\phi} \right) = \frac{1}{2} \hat{\bm{\phi}} \cdot \left( \mathscr{D}^{\mu} [\mathscr{A}] \mathscr{D}_{\mu} [\mathscr{A}] \hat{\bm{\phi}} \right)
,
\end{equation}
which is used to eliminate the Lagrange multiplier field $u$ in (\ref{YM_eq_scalar}). 
Indeed, the field equations (\ref{YM_eq_gauge}) and (\ref{YM_eq_scalar}) are rewritten in terms of $\mathscr{A}_{\mu} (x)$ and $\hat{\bm{\phi}} (x)$ into
\begin{align}
\mathscr{D}^{\mu} [\mathscr{A}] \mathscr{F}_{\mu \nu} - g v^{2} \hat{\bm{\phi}} \times  \mathscr{D}_{\nu} [\mathscr{A}] \hat{\bm{\phi}} = & 0 , 
\label{eq_gauge1} \\
\mathscr{D}^{\mu} [\mathscr{A}] \mathscr{D}_{\mu} [\mathscr{A}] \hat{\bm{\phi}}  - \left( \hat{\bm{\phi}} \cdot  \mathscr{D}^{\mu} [\mathscr{A}] \mathscr{D}_{\mu} [\mathscr{A}] \hat{\bm{\phi}} \right) \hat{\bm{\phi}} = & 0
\label{eq_scalar1}
.
\end{align}
By applying the covariant derivative $\mathscr{D}^{\nu} [\mathscr{A}]$ to the equation (\ref{eq_gauge1}), the reduction condition is naturally induced:
\begin{align}
0 = & \mathscr{D}^{\nu} [\mathscr{A}] \mathscr{D}^{\mu} [\mathscr{A}] \mathscr{F}_{\mu \nu} 
= g v^{2} \hat{\bm{\phi}} \times \mathscr{D}^{\nu} [\mathscr{A}] \mathscr{D}_{\nu} [\mathscr{A}] \hat{\bm{\phi}} = g v^{2} \bm{\chi}
.
\end{align}
Moreover,  by taking the exterior product of (\ref{eq_scalar1}) and $\hat{\bm{\phi}} (x)$, the reduction condition is induced again:
\begin{align}
0 = & \hat{\bm{\phi}} \times \mathscr{D}^{\mu} [\mathscr{A}] \mathscr{D}_{\mu} [\mathscr{A}] \hat{\bm{\phi}} - \left( \hat{\bm{\phi}} \cdot \mathscr{D}^{\mu} [\mathscr{A}] \mathscr{D}_{\mu} [\mathscr{A}] \hat{\bm{\phi}} \right) \left( \hat{\bm{\phi}} \times \hat{\bm{\phi}} \right) \nonumber\\
= & \hat{\bm{\phi}} \times \mathscr{D}^{\mu} [\mathscr{A}] \mathscr{D}_{\mu} [\mathscr{A}] \hat{\bm{\phi}} = \bm{\chi}
.
\end{align}
Hence, the simultaneous solutions of the coupled field equations (\ref{eq_gauge1}) and (\ref{eq_scalar1}) automatically satisfy the reduction condition (\ref{reduction2}).
From this relation, we find that the solutions of the coupled field equations of the gauge-scalar model (\ref{eq_gauge1}) and (\ref{eq_scalar1}) can become the field configurations satisfying the reduction condition (\ref{reduction2}), which gives the field configuration to be taken into account in constructing the massive Yang--Mills theory through the path-integral (\ref{path_integral}).

\section{Julia--Zee dyon solution in the Georgi--Glashow model}

In this section, we give a brief review of the Julia--Zee dyon \cite{Julia--Zee} in the ($3+1$)-dimensional Minkowski spacetime $\mathbb{R}^{1,3}$. 
The Georgi--Glashow model is introduced by the Lagrangian density
\begin{equation}
\mathscr{L}_{\rm GG} = - \frac{1}{4} \mathscr{F}_{\mu \nu} \cdot \mathscr{F}^{\mu \nu} + \frac{1}{2} \left( \mathscr{D}_{\mu} [\mathscr{A}] \bm{\phi} \right) \cdot \left( \mathscr{D}^{\mu} [\mathscr{A}] \bm{\phi} \right) - \frac{\lambda^{2} g^{2}}{4} \left( \bm{\phi} \cdot \bm{\phi} - v^{2} \right)^{2}
,
\end{equation}
where $g, \lambda$ and $v > 0$ are respectively the gauge coupling constant, the scalar coupling constant and the value of the magnitude $|\bm{\phi} (x) |$ of the adjoint scalar field $\bm{\phi} (x)$ at the vacuum which is to be realized at infinity $|x| = \infty$. 

By varying the action
\begin{equation}
S = \int d^{4} x \ \mathscr{L}_{\rm GG}
,
\end{equation}
with respect to the fields $\mathscr{A}_{\mu} (x)$ and $\bm{\phi} (x)$, the field equations are obtained as
\begin{align}
& \mathscr{D}^{\mu} [\mathscr{A}] \mathscr{F}_{\mu \nu} - g \bm{\phi} \times \mathscr{D}_{\nu} [\mathscr{A}] \bm{\phi} =  0 , \label{eq_A}\\
& \mathscr{D}^{\mu} [\mathscr{A}] \mathscr{D}_{\mu} [\mathscr{A}] \bm{\phi} + \lambda^{2} g^{2} \left( \bm{\phi} \cdot \bm{\phi} - v^{2} \right) \bm{\phi} =  0
.\label{eq_phi}
\end{align}

The Julia--Zee ansatz with a unit magnetic charge is given by
\begin{equation}
g \mathscr{A}_{0}^{A} (x) = \frac{x^{A}}{r} \widetilde{a} (r) , \ \ \ 
g \mathscr{A}_{j}^{A} (x) = \epsilon^{j A k} \frac{x^{k}}{r} \frac{1 - \widetilde{f} (r)}{r} , \ \ \ 
\phi^{A} (x) = v \frac{x^{A}}{r} \widetilde{h} (r)
,
\label{Julia--Zee}
\end{equation}
where Roman indices $j,k$ run from $1$ to $3$ and $r$ is the radius $r : = \sqrt{x^{2} + y^{2}+ z^{2}}$ in the three-dimensional space with the Cartesian coordinates $(x, y ,z)$.
Note that the electric charge cannot be specified at this stage in contrast to the magnetic charge specified by the form of the ansatz (\ref{Julia--Zee}) for $\mathscr{A}_{j}^{A} (x)$ which has the same form as that used for obtaining the 't Hooft--Polyakov magnetic monopole with a unit magnetic charge \cite{tHP}.
This is because the electric charge depends on the asymptotic value of the profile function $\widetilde{a} (r)$ in $\mathscr{A}_{0}^{A} (x)$, to be obtained by solving the coupled field equations simultaneously for the other unknown functions $\widetilde{f} (r)$ and $\widetilde{h} (r)$, as will be performed below.


The field equations (\ref{eq_A}) and (\ref{eq_phi}) are rewritten in terms of the profile functions $\widetilde{a}$, $\widetilde{f}$, and $\widetilde{h}$  as 
\begin{align}
& \widetilde{a}^{\prime \prime} (r) + \frac{2}{r} \widetilde{a}^{\prime} (r) - \frac{2}{r^{2}} \widetilde{a} (r) \widetilde{f}^{2} (r) = 0 , \label{GG_eq_1}\\
& \widetilde{f}^{\prime \prime} (r) - \frac{\widetilde{f}^{3} (r) - \widetilde{f} (r)}{r^{2}} + \left( \widetilde{a}^{2} (r) - g^{2} v^{2} \widetilde{h}^{2} (r) \right) \widetilde{f} (r) = 0 , \label{GG_eq_2}\\
& \widetilde{h}^{\prime \prime} (r) + \frac{2}{r} \widetilde{h}^{\prime} (r) - \frac{2}{r^{2}} \widetilde{h} (r) \widetilde{f}^{2} (r) - \lambda^{2} g^{2} v^{2} \left( \widetilde{h}^{3} (r) - \widetilde{h} (r) \right) = 0
.\label{GG_eq_3}
\end{align}
The scaled dimensionless variable $\rho$, and the scaled functions $a$, $f$, and $h$ of $\rho$ defined by
\begin{equation}
\rho := g v r , \ \ \ 
\widetilde{a} (r) := g v a (\rho) , \ \ \ 
\widetilde{f} (r) = f (\rho) , \ \ \ 
\widetilde{h} (r) = h (\rho) 
,
\label{dimensionless}
\end{equation}
are introduced to make the field equations (\ref{GG_eq_1})--(\ref{GG_eq_3}) dimensionless:
\begin{align}
& a^{\prime \prime} (\rho) + \frac{2}{\rho} a^{\prime} (\rho) - \frac{2}{\rho^{2}} a (\rho) f^{2} (\rho) = 0 , 
\label{GG_eq_a_dyon} \\
& f^{\prime \prime} (\rho) - \frac{f^{3} (\rho) - f (\rho)}{\rho^{2}} + \left( a^{2} (\rho) - h^{2} (\rho) \right) f (\rho) = 0 , 
\label{GG_eq_f_dyon} \\
& h^{\prime \prime} (\rho) + \frac{2}{\rho} h^{\prime} (\rho) - \frac{2}{\rho^{2}} h (\rho) f^{2} (\rho)  - \lambda^{2} ( h^{3} (\rho) - h (\rho) ) = 0
\label{GG_eq_h_dyon} ,
\end{align}
where the prime denotes the derivative with respect to $\rho$ hereafter.

In order to determine the boundary conditions, we consider the static energy $E$ given by
\begin{align}
E = \frac{4 \pi M_{\mathscr{X}}}{g^{2}} 
\int_{0}^{\infty} d \rho \ e (\rho)
,
\label{dyon_energy}
\end{align}
where we have defined the mass scale $M_{\mathscr{X}}$ by
\begin{equation}
M_{\mathscr{X}} := g v
,
\end{equation}
and the energy density $e (\rho)$ by
\begin{align}
e (\rho) := & \frac{1}{2} \rho^{2} a^{\prime 2} (\rho) + a^{2} (\rho) f^{2} (\rho) + f^{\prime 2} (\rho) + \frac{( f^{2} (\rho) - 1)^{2}}{2 \rho^{2}} \nonumber\\
& +  \frac{1}{2} \rho^{2} h^{\prime 2} (\rho) + h^{2} (\rho) f^{2} (\rho) + \frac{\lambda^{2} }{4} \rho^{2} \left( h^{2} (\rho) - 1 \right)^{2}
.\label{energy_density}
\end{align}

For the energy (\ref{dyon_energy}) to be finite, the regularity of the fields at $\rho = 0$ is required:
\begin{equation}
a (0) = 0 , \ \ \ 
f (0) = 1  , \ \ \ 
h (0) = 0
.
\label{boundary_0}
\end{equation}

For large $\rho$, by the same reason, the scalar field must go to its vacuum expectation value at infinity:
\begin{equation}
| \bm{\phi} (x) | \xrightarrow{|x| \to \infty} v \ \ \ \Rightarrow \ \ \ h (\infty) = 1
.
\end{equation}
As the gauge field $\mathscr{A}_{\mu}^{A} (x)$ goes to the pure gauge form at $\rho \to \infty$, the profile function $f (\rho)$ should take
\begin{equation}
f (\infty) = 0
.
\label{f_bc}
\end{equation}
In order to solve the field equations (\ref{GG_eq_a_dyon})--(\ref{GG_eq_h_dyon}), however, the asymptotic value $a_{\infty}$ of $a (\rho)$ must be specified: 
\begin{equation}
 a_{\infty} := a(\infty).
\end{equation}
Notice that $a_{\infty}$ is not completely arbitrary.
As will be shown below, indeed, for $f (\rho)$ not to oscillate at large $\rho$ so that the spatial components $\mathscr{A}_{j}^{A} (x)$ of the gauge field become the pure gauge form at $\rho \approx \infty$, the constant $a_{\infty}$ should take the value $| a_{\infty} | < 1$.  
Notice that, if $a(\rho)$ is a solution of equations (\ref{GG_eq_a_dyon})--(\ref{GG_eq_h_dyon}), then $-a(\rho)$ is also a solution of them. 
Therefore, $a_{\infty}$ is restricted to take the nonnegative value $0 \leq a_{\infty} < 1$ without loosing the generality.  
The solution  $a (\rho) \equiv 0$ with a vanishing $\mathscr{A}_{0}$ component corresponds to the 't Hooft--Polyakov magnetic monopole, which leads to $a_{\infty} = 0$.
Since the parameters $g$ and $v$ are factorized out, dyon solutions are distinguished by the value of $a_{\infty}$ and the scalar coupling $\lambda$.

We further consider the asymptotic forms of the profile functions.
For small $\rho$, $\rho \approx 0$, we assume the power-series expansion in $\rho$
\begin{equation}
a (\rho) = \sum_{n = 1}^{\infty} A_{n} \rho^{n} , \ \ \ 
f (\rho) = 1 + \sum_{n = 1}^{\infty} F_{n} \rho^{n} , \ \ \ 
h (\rho) = \sum_{n = 1}^{\infty} H_{n} \rho^{n} 
.
\end{equation}
By substituting the above power-series expansion into the field equations (\ref{GG_eq_a_dyon})--(\ref{GG_eq_h_dyon}), the asymptotic forms for small $\rho$ are obtained
\begin{align}
a (\rho) \approx & A_{1} \biggl[ \rho + \frac{2}{5} F_{2} \rho^{3} + \frac{12 F_{2}^{2} + H_{1}^{2} - A_{1}^{2}}{70} \rho^{5} + \cdots \biggr] , \\
f (\rho) \approx & 1 + F_{2} \rho^{2} + \frac{3 F_{2}^{2} + H_{1}^{2} - A_{1}^{2}}{10} \rho^{4} + \frac{\lambda^{2} H_{1}^{2} + 12 F_{2} \left( H_{1}^{2} - A_{1}^{2} \right) + 14 F_{2}^{3}}{140} \rho^{6} + \cdots , \\
h (\rho) \approx & H_{1} \biggl[ \rho + \frac{\lambda^{2} + 4 F_{2}}{10} \rho^{3} + \frac{\lambda^{4} - 10 \lambda^{2} H_{1}^{2} + 4 H_{1}^{2} + 8 \lambda^{2} F_{2} + 48 F_{2}^{2} - 4 A_{1}^{2}}{280} \rho^{5} + \cdots \biggr]
,
\end{align}
in agreement with the boundary conditions (\ref{boundary_0}) for small $\rho$.

For large $\rho$, $\rho \to \infty$, by introducing $b (\rho)$ and $k (\rho)$ by
\begin{equation}
a (\rho) := a_{\infty} + b (\rho) , \ \ \ 
h (\rho) := 1 + k (\rho)
,
\end{equation}
the field equations (\ref{GG_eq_a_dyon})--(\ref{GG_eq_h_dyon}) reduce to the linear differential equations
\begin{align}
& b^{\prime \prime} (\rho) + \frac{2}{\rho} b^{\prime} (\rho) \approx 0 , \label{asymp_a} \\
& f^{\prime \prime} (\rho) + \left( a_{\infty}^{2} - 1 \right) f (\rho) \approx 0 , \label{asymp_f} \\
& k^{\prime \prime} (\rho) + \frac{2}{\rho} k^{\prime} (\rho) - 2 \lambda^{2} k (\rho) \approx 0 
.
\label{asymp_k}
\end{align}
These equations can be solved independently under the boundary conditions $b (\infty) = 0$, $f (\infty) = 0$, and $k (\infty) = 0$ as
\begin{align}
b (\rho) \approx & - \frac{C}{\rho}  , 
\label{JZ_dyon_asymp_a} \\
f (\rho) \approx &  F \exp \left\{ - \sqrt{1 - a_{\infty}^{2}} \rho \right\}
,\label{JZ_dyon_asymp_f} \\
k (\rho) \approx & H \frac{e^{- \sqrt{2} \lambda \rho}}{\rho} 
,
\label{JZ_dyon_asymp_h}
\end{align}
where $C, F$ and $H$ are arbitrary constants.
The asymptotic solution (\ref{JZ_dyon_asymp_f}) for large $\rho$ indicates that for the profile function $f (\rho)$ of the spatial components $\mathscr{A}_{j} (x)$ of the gauge field not to oscillate at large $\rho$, the constant $a_{\infty}$ should take the value $|a_{\infty}| < 1$. Therefore, $a_{\infty}$ can be restricted to $0 \leq a_{\infty} < 1$ without loosing the generality.

We define the chromomagnetic field $\mathscr{B}_{j}^{A} (x)$ and chromoelectric field $\mathscr{E}_{j}^{A} (x)$ by
\begin{equation}
\mathscr{B}_{j}^{A} (x) := \frac{1}{2} \epsilon_{j k l} \mathscr{F}_{k l}^{A} (x) , \ \ \ 
\mathscr{E}_{j}^{A} (x) := \mathscr{F}_{j 0}^{A} (x)
,
\end{equation}
and the magnetic charge $q_{m}$ and electric charge $q_{e}$ by
\begin{align}
q_{m} := & \int d^{3} x \ \mathscr{B}_{j} \cdot \left( \mathscr{D}_{j} [\mathscr{A}] \hat{\bm{\phi}} \right) , \\
q_{e} := & \int d^{3} x \ \mathscr{E}_{j} \cdot \left( \mathscr{D}_{j} [\mathscr{A}] \hat{\bm{\phi}} \right)
,
\end{align}
where we have introduced the normalized scalar field 
\begin{equation}
\hat{\bm{\phi}} (x) := \frac{1}{v} \bm{\phi} (x)
.
\end{equation}
By using the asymptotic forms (\ref{JZ_dyon_asymp_a})--(\ref{JZ_dyon_asymp_f}),  the magnetic and electric charges $q_{m}$ and $q_{e}$ are calculated as
\begin{align}
q_{m} := &
 \frac{4 \pi}{g} \int_{0}^{\infty} d \rho \ \frac{d}{d \rho} \bigl[ h (\rho) \left( 1 - f^{2} (\rho) \right) \bigr] \nonumber\\
=& \frac{4 \pi}{g} \bigl[ h(\rho) \left( 1 - f^{2} (\rho) \right) \bigr] \biggl|_{\rho=0}^{\rho = \infty} =  \frac{4 \pi}{g} , \\
q_{e} := &  \int_{S^{2}} d^{2} S_{j} \ \mathscr{E}_{j} \cdot \hat{\bm{\phi}} \nonumber\\
= & \lim_{\rho \to \infty} \frac{4 \pi}{g} \rho^{2} \frac{x^{j} x^{A}}{\rho^{2}} \biggl[ \frac{x^{j} x^{A}}{\rho^{2}} a^{\prime} (\rho) + \left( \frac{\delta^{A j}}{\rho} - \frac{x^{A} x^{j}}{\rho^{3}} \right) a (\rho) f (\rho) \biggr] \nonumber\\
= & \frac{4 \pi}{g} \lim_{\rho \to \infty} \rho^{2} \left( \frac{C}{\rho^{2}} \right) = \frac{4 \pi}{g} C = q_{m} C 
,
\label{electric_charge}
\end{align}
where we assume that the coordinates $x^{j}$ are also dimensionless as well as $\rho$.
We find that the magnetic charge $q_{m}$ is indeed nontrivial and has a minimal value in unit of $4 \pi /g$.
We also find that the coefficient $C$ of the next-leading term of $a(\rho)$ in (\ref{JZ_dyon_asymp_a}) is nothing but the ratio of the charges $q_{e} / q_{m}$:
\begin{equation}
a (\rho) = a_{\infty} - \frac{C}{\rho} + \cdots, \ \  \ C = \frac{q_{e}}{q_{m}}
.
\end{equation}
It should be noticed that, although some physical quantities such as the chromoelectric field $\mathscr{E}_{j}^{A} (x)$ and the electric charge $q_{e}$ do not depend on the asymptotic value $a_{\infty}$ of $a (\rho)$, they depend on the next-leading coefficient $C$ of $a (\rho)$, namely, the ratio of the charges. 
Therefore, in order to compare the solutions and the corresponding physical quantities with the same electric charge $q_{e}$ by varying the scalar coupling $\lambda$, we adopt the following boundary condition 
\begin{equation}
\rho^{2} a^{\prime} (\rho) \xrightarrow{\rho \to \infty} C = \frac{q_{e}}{q_{m}}
.
\end{equation}
If we restrict the solution $a(\rho)$ to non-negative one $a(\rho)>0$ as mentioned in the above, then we have only $a_{\infty}\geq0$ and $C=q_{e}/q_{m}\geq0$.

\section{Construction of the Yang--Mills dyon}

Next, we discuss the dyon in the massive $SU(2)$ Yang--Mills theory through the ``complementary'' $SU(2)$ gauge-scalar model.
Taking the Julia--Zee ansatz (\ref{Julia--Zee}), the field equations (\ref{YM_eq_gauge}), (\ref{YM_eq_scalar}), and the constraint (\ref{constraint}) become
\begin{align}
& \widetilde{a}^{\prime \prime} (r) + \frac{2}{r} \widetilde{a}^{\prime} (r) - \frac{2}{r^{2}} \widetilde{a} (r) \widetilde{f}^{2} (r) = 0 , \\
& \widetilde{f}^{\prime \prime} (r) - \frac{\widetilde{f}^{3} (r) - \widetilde{f} (r)}{r^{2}} + \left( \widetilde{a}^{2} (r) - g^{2} v^{2} \widetilde{h}^{2} (r) \right) \widetilde{f} (r) = 0 , \\
& \widetilde{h}^{\prime \prime } (r) + \frac{2}{r} \widetilde{h}^{\prime} (r) - \frac{2}{r^{2}} \widetilde{h} (r) \widetilde{f}^{2} (r) +2  u (r)  \widetilde{h} (r) = 0 , \\
& \widetilde{h}^{2} (r) = 1
.
\end{align}
Here the last equation is nothing but the radially fixing constraint and can be used to eliminate $\widetilde{h} (r)$ from the other equations to obtain
\begin{align}
& \widetilde{a}^{\prime \prime} (r) + \frac{2}{r} \widetilde{a}^{\prime} (r) - \frac{2}{r^{2}} \widetilde{a} (r) \widetilde{f}^{2} (r) = 0 , 
\label{YM_eq_a}\\
& \widetilde{f}^{\prime \prime} (r) - \frac{\widetilde{f}^{3} (r) - \widetilde{f} (r)}{r^{2}} + \left( \widetilde{a}^{2} (r) - g^{2} v^{2}  \right) \widetilde{f} (r) = 0 , 
\label{YM_eq_f}\\
& u (r) =  \frac{1}{r^{2}} \widetilde{f}^{2} (r) 
\label{determine_u}
.
\end{align}
Hence, the Lagrange multiplier field $u = u (r)$ can be determined through (\ref{determine_u}) once the remaining two equations (\ref{YM_eq_a}) and (\ref{YM_eq_f}) for $\widetilde{a} (r)$ and $\widetilde{f} (r)$ are solved.

In order to make the field equations dimensionless, we define the dimensionless variable $\rho = g v r$ and the rescaled functions of $\rho$: $\widetilde{a} (r) = g v a (\rho)$, $\widetilde{f} (r) = f (\rho)$.
Then (\ref{YM_eq_a}) and (\ref{YM_eq_f}) read
\begin{align}
& a^{\prime \prime} (\rho) + \frac{2}{\rho} a^{\prime} (\rho) - \frac{2}{\rho^{2}} a (\rho) f^{2} (\rho) = 0 , 
\label{YM_dyon_eq_a}\\
& f^{\prime \prime} (\rho) - \frac{f^{3} (\rho) - f (\rho)}{\rho^{2}} + \left( a^{2} (\rho) - 1 \right) f (\rho) = 0 
\label{YM_dyon_eq_f}
.
\end{align}
By repeating the same procedure for obtaining the boundary condition as the Julia--Zee dyon, 
we find that it is sufficient to impose the following boundary conditions for the Yang--Mills dyon:
\begin{align}
 a (0) = & 0 , \ \ \ \ \ \ f (0) = 1, 
\label{YM_dyon_boundary_0} \\
 \rho^{2} a^{\prime} (\rho) \xrightarrow{\rho \to \infty} & C, \ \ \ \ \ \ f (\infty) = 0 
,
\end{align}
where $C$ is an arbitrary constant.
We find  that these conditions are enough to guarantee the regularity of the fields at the origin $\rho = 0$ to obtain a finite energy.

\begin{figure}[t]
\centering
\includegraphics[width=0.45\textwidth]{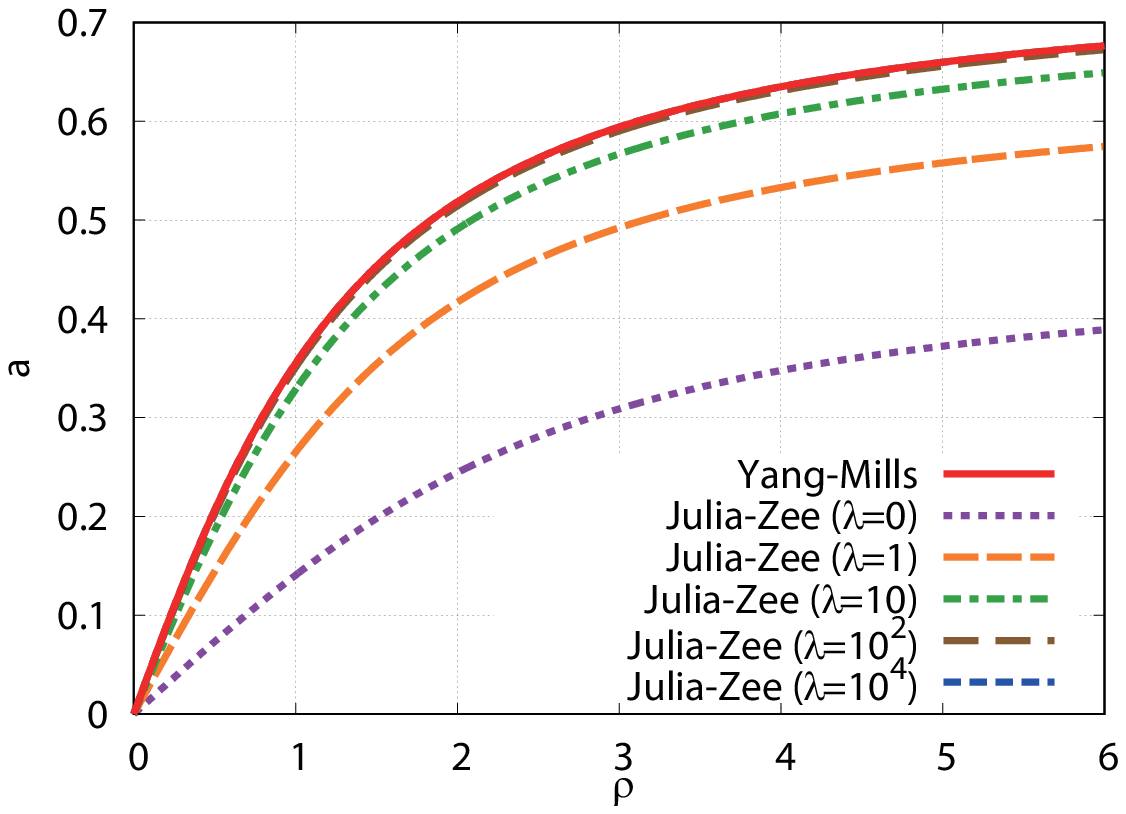} \ 
\includegraphics[width=0.45\textwidth]{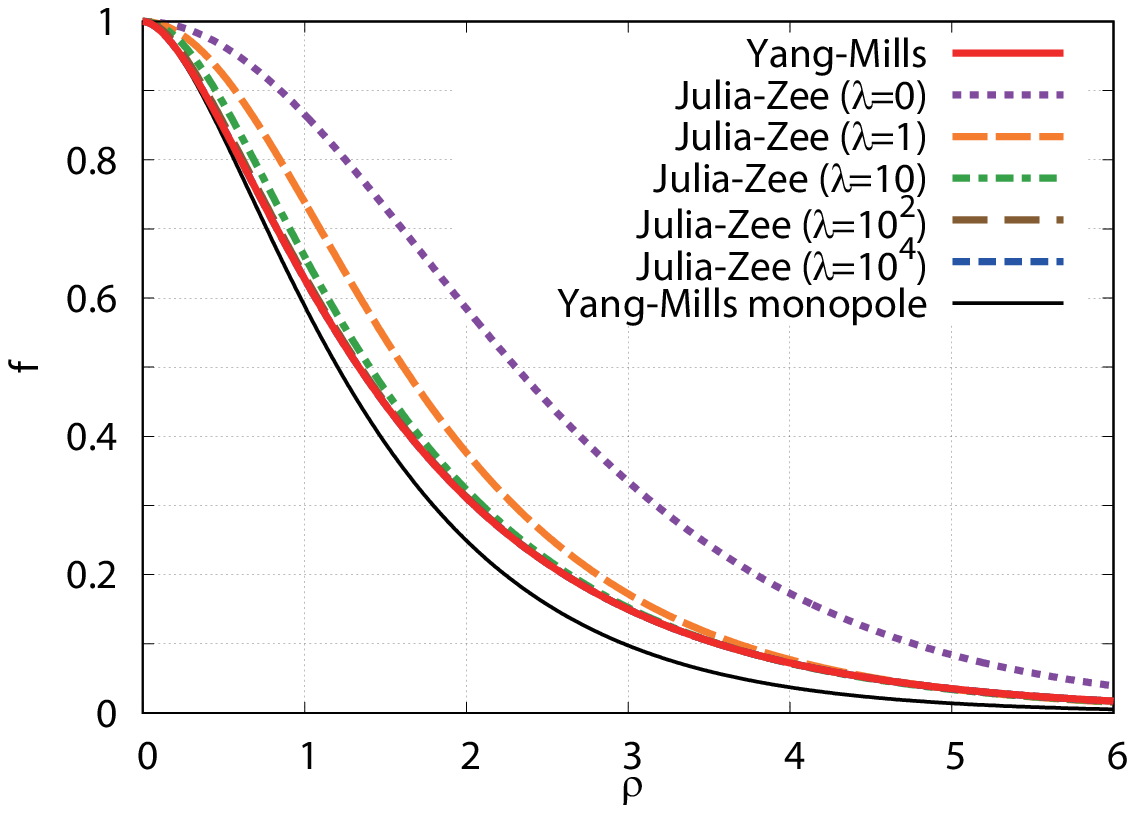} \\ 
\includegraphics[width=0.45\textwidth]{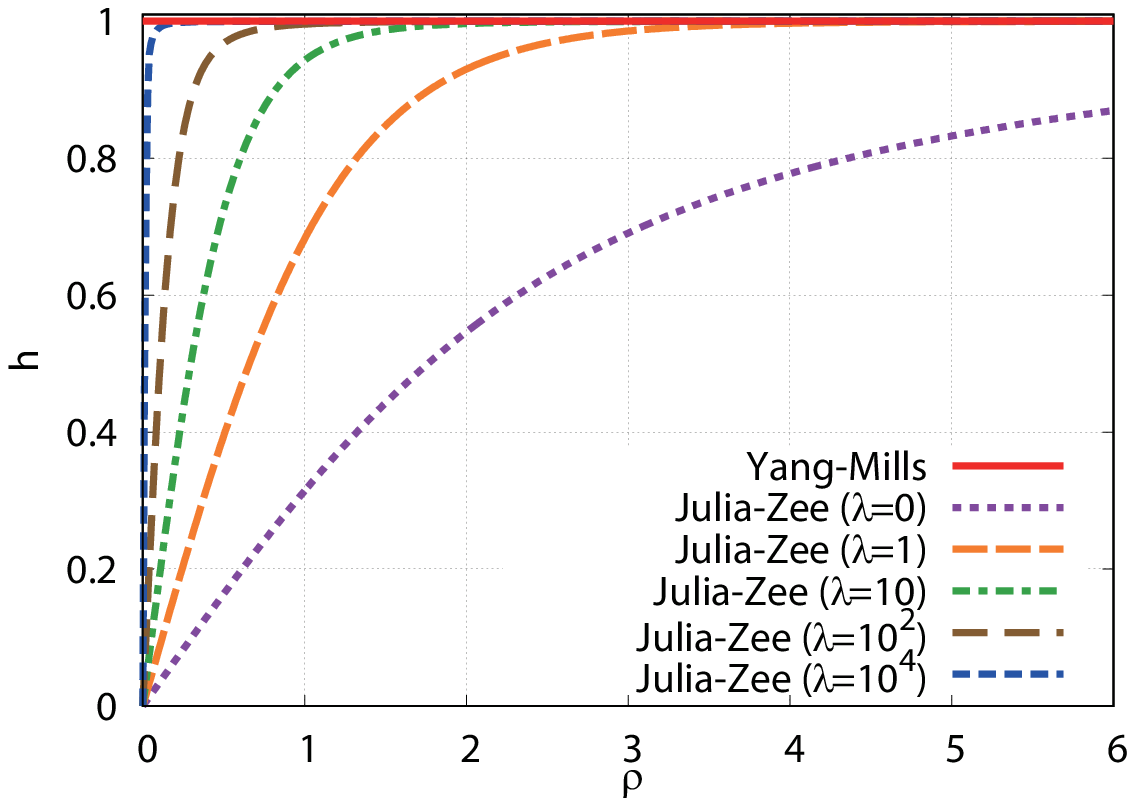} \ 
\includegraphics[width=0.45\textwidth]{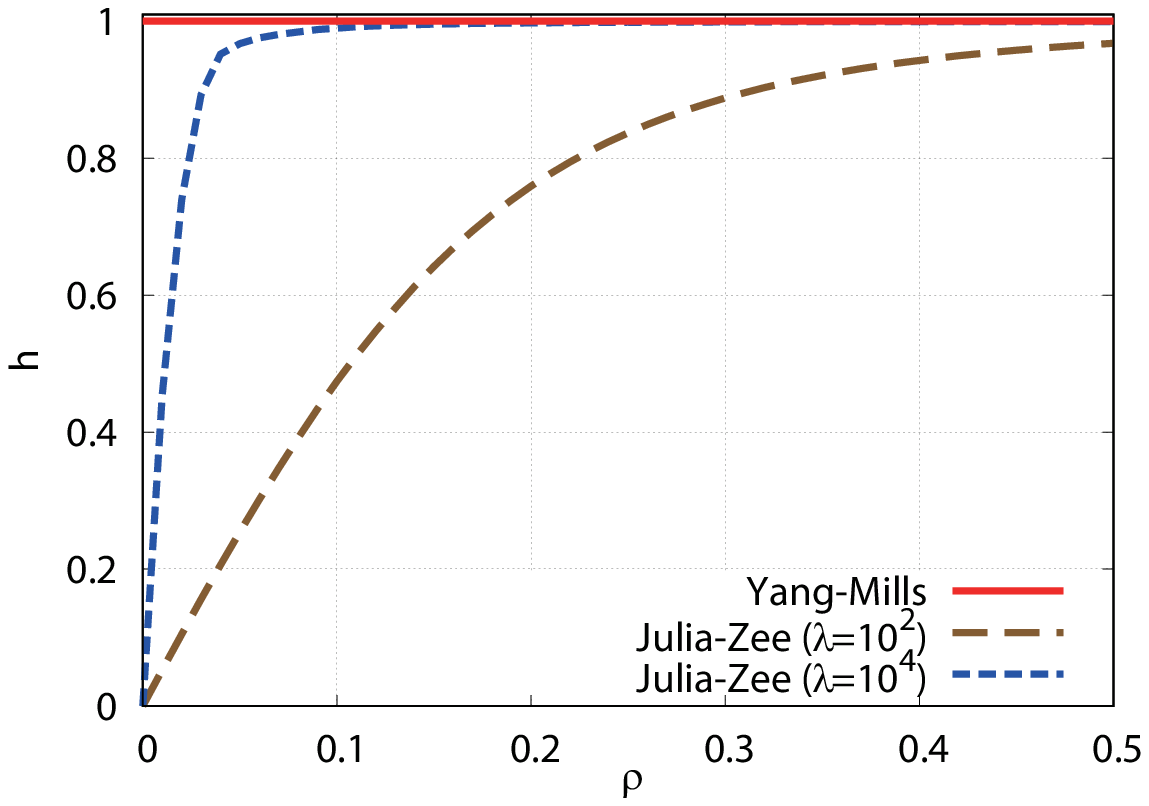}
\caption{(Top) The solutions $a$ (left panel) and $f$ (right panel) of the Yang--Mills dyon equations (\ref{YM_dyon_eq_a}) and (\ref{YM_dyon_eq_f}) as functions of $\rho = g v r$, which are to be compared with the Yang--Mills monopole and the Julia--Zee dyon solutions for $\lambda = 0, 1, 10, 10^{2}$, and $10^{4}$ for $C=0.5$.
(Bottom) The corresponding solution $h$ for the scalar field as a function of $\rho$.
The radially fixing constraint $h (\rho) \equiv 1$ holds even at the origin $\rho = 0$ in the Yang--Mills dyon, while the naive $\lambda \to \infty$ limit of the Julia--Zee dyon approaches the limit value, $h_{\rm JZ} (\rho) \to 1$ only for $\rho >0$ except the origin $\rho = 0$.
}
\label{dyon_solution}
\end{figure}

For large $\rho$, the equations (\ref{YM_dyon_eq_a}) and (\ref{YM_dyon_eq_f}) reduce to (\ref{asymp_a}) and (\ref{asymp_f}) respectively and therefore the profile functions behave like the Julia--Zee dyon:
\begin{equation}
a (\rho) \approx a_{\infty} - \frac{C}{\rho} , \ \ \ f (\rho) \approx F \exp \left\{ - \sqrt{1 - a_{\infty}^{2}} \rho \right\}
.
\label{YM_dyon_asymptotic_form}
\end{equation}

For small $\rho$, however, the asymptotic forms are much different from the Julia--Zee case.
To realize this, we shall linearize the field equations by assuming $f (\rho) = 1 + g (\rho)$ and $|g (\rho) | , |a (\rho)| \ll 1$.
Then, the field equations (\ref{YM_dyon_eq_a}) and (\ref{YM_dyon_eq_f}) become
\begin{align}
& \rho^{2} a^{\prime \prime} (\rho) + 2 \rho a^{\prime} (\rho) - 2 a (\rho) = 0 , \label{YM_asymp_a}\\
& \rho^{2} g^{\prime \prime} (\rho) - 2 g (\rho) - \rho^{2} g (\rho) = \rho^{2} 
.\label{YM_asymp_g}
\end{align}
The first equation (\ref{YM_asymp_a}) is solved as
\begin{equation}
a (\rho) \approx A_{1} \rho \ \ \ \ (\rho \approx 0)
 .
\end{equation}
The second equation (\ref{YM_asymp_g}) is the same as the Yang--Mills monopole case and can be solved as \cite{Nishino}
\begin{equation}
f (\rho) - 1 = g (\rho) \approx \widetilde{C}^{2} \rho^{2} + \frac{1}{3} \rho^{2} \log \rho + \cdots \ \ \ \ (\rho \approx 0)
.
\end{equation}
Here note that the equation cannot be satisfied by a simple power-series of $\rho$ without the logarithmic term.

The field equations (\ref{YM_dyon_eq_a}) and (\ref{YM_dyon_eq_f}) can be solved numerically, which is shown in Fig. \ref{dyon_solution} for e.g., $C = 0.5$.
The Julia--Zee dyon solution with a large coupling $\lambda \gg 1$ approaches the Yang--Mills dyon except for the neighborhood of the origin $\rho \approx 0$.
This is the similar situation to the Yang--Mills monopole.
For $a (\rho)$, the naive limit $\lambda \to \infty$ of the Julia--Zee dyon completely agrees with the Yang--Mills dyon.

In particular, $f (\rho) \equiv 0$ is also the solution of (\ref{YM_dyon_eq_f}) and hence the exact solution of (\ref{YM_dyon_eq_f})  is given by 
\begin{equation}
a (\rho) = a_{\infty} - \frac{C}{\rho}
.
\label{WY_dyon}
\end{equation}
These solutions, however, do not satisfy the boundary conditions (\ref{YM_dyon_boundary_0}) for $\rho \to 0$ and $a (\rho)$ diverges at the origin $\rho = 0$ for $C \neq 0$.
In view of these, the dyon constructed by $f (\rho) \equiv 0$ and (\ref{WY_dyon}) has a diverging energy and is regarded as a dyonic extension of the Wu--Yang monopole.
For $C = 0$, the solution leads $a (\rho) \equiv 0$, which means that this solution is nothing but the Wu--Yang monopole with a vanishing electric charge.

\section{Gauge field decomposition  for a dyon}

In what follows, we shall omit the tilde $(\tilde{})$ for the profile functions $\widetilde{f}$ and $\widetilde{h}$ to simplify the notation.

In the present ansatz (\ref{Julia--Zee}), the normalized scalar field $\hat{\bm{\phi}} (x)$ with $h (r) = +1$,
\begin{equation}
\hat{\phi}^{A} (x) = \frac{x^{A}}{r}
,
\end{equation}
leads to the decomposed fields which are explicitly written as
\begin{align}
g \mathscr{V}^{A}_{0} (x) = &  \frac{x^{A}}{r} \widetilde{a} (r) , \ \ \ g \mathscr{X}^{A}_{0} (x) = 0 , \\
g \mathscr{V}^{A}_{j} (x) = & \frac{\epsilon^{j A k} x^{k}}{r} \frac{1}{r}  , \ \ \ 
g \mathscr{X}^{A}_{j} (x) = - \frac{\epsilon^{j A k} x^{k}}{r} \frac{f (r)}{r}
.
\end{align}
Notice that the time component of $\mathscr{X}_\mu(x)$ is identically zero $\mathscr{X}_0(x)=0$, according to (\ref{decomposition}) by taking into account the facts that $\hat{\bm{\phi}} (x)$ is time-independent and that the time component $\mathscr{A}_{0} (x)$ of the gauge field and the normalized scalar field $\hat{\bm{\phi}} (x)$ are parallel in the color space $g \mathscr{A}^{A}_{0} (x) = \widetilde{a} (r) \hat{\phi}^{A} (x)$:  
\begin{equation}
g \mathscr{X}_{0} (x) = \hat{\bm{\phi}} (x) \times \left( \partial_{0} \hat{\bm{\phi}} (x) - g \mathscr{A}_{0} (x) \times \hat{\bm{\phi}} (x) \right) = 0
.
\end{equation}

In what follows, we adopt the polar coordinate system $(r , \theta , \varphi)$ for the spatial coordinates. 
Then the decomposed field have the following components:
\begin{align}
g \mathscr{A}_{0} (x) = & A_{0} (r) T_{r} , \ \ \ 
g \mathscr{A}_{r} (x) = 0 , \ \ \ 
g \mathscr{A}_{\theta} (x) = A (r) T_{\theta} , \ \ \ 
g \mathscr{A}_{\varphi} (x) = A (r) T_{\varphi} , \\
g \mathscr{V}_{0} (x) = & V_{0} (r) T_{r} , \ \ \  \ 
g \mathscr{V}_{r} (x) = 0 , \ \ \ 
g \mathscr{V}_{\theta} (x) =V (r) T_{\theta} , \ \ \ \ 
g \mathscr{V}_{\varphi} (x) = V (r) T_{\varphi} , \\
g \mathscr{X}_{0} (x) = & 0 , \ \ \ \ \ \ \ \ \ \ \ \ 
g \mathscr{X}_{r} (x) = 0 , \ \ \ 
g \mathscr{X}_{\theta} (x) = X (r) T_{\theta} , \ \ \ 
g \mathscr{X}_{\varphi} (x) =X (r) T_{\varphi}
,
\end{align}
and
\begin{equation}
A_{0} (r) = V_{0} (r) = \widetilde{a} (r) , \ \ \ 
A (r) = \frac{1 - f ( r)}{r} , \ \ \ 
V (r) = \frac{1}{r} , \ \ \ 
X (r) = - \frac{f (r)}{r}
 ,
\end{equation}
where we have defined
\begin{align}
T_{r} = \frac{1}{2} \begin{pmatrix}
\cos \theta & \sin \theta e^{- i \varphi} \\
\sin \theta e^{i \varphi} & - \cos \theta
\end{pmatrix}
, \   
T_{\theta} = \frac{1}{2} \begin{pmatrix}
0 & -i e^{- i \varphi} \\
i e^{i \varphi} & 0 
\end{pmatrix}
, \   
T_{\varphi} = \frac{1}{2} \begin{pmatrix}
 \sin \theta & -\cos \theta e^{- i \varphi} \\
-\cos \theta e^{i \varphi}  & -\sin \theta
\end{pmatrix}
.
\end{align}

\begin{figure}[t]
\centering
\includegraphics[width=0.45\textwidth]{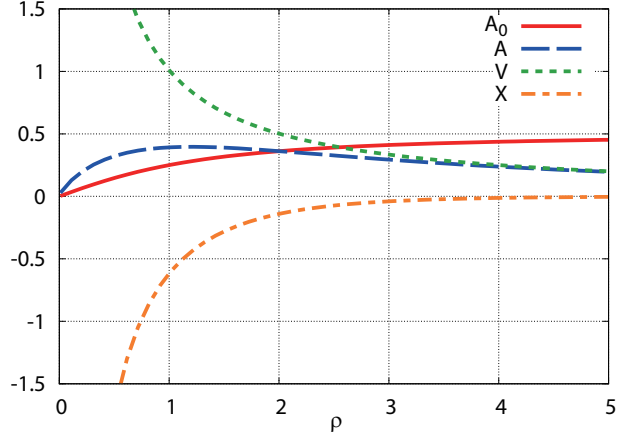}
\caption{The fields of $A_{0}, A, V$, and $X$ as functions of $\rho = g v r$ at $a_{\infty} = 0.5$.
Here $A (x) = V (x) + X (x)$ where $V (x)$ agrees with the Wu--Yang monopole and $X(x)$ corresponds to the massive mode.
The time component $A_{0} (x)$ is finite by itself.}
\label{dyon_fields}
\end{figure}

Fig. \ref{dyon_fields} is the plot of the fields $A_{0}$, $A$ ,$V$, and $X$ as functions of $\rho = g v r$.
We find that the spatial component $A(r)$ of the original gauge field $\mathscr{A}_\mu(x)$ is indeed regular at the origin: 
\begin{equation}
A (r) \approx g v \biggl[  \widetilde{C}^{2} r + \frac{1}{3} r \log \left( g v r \right) \biggr] \ \ \ (r \approx 0)
.
\end{equation}
It should be remarked that although both decomposed fields $V(x)$ and $X(x)$ are singular at the origin, the singularities cancel between the two decomposed fields to yield the regular $A(r)$. 
On the other hand, the time component $A_{0} (r)$ of the gauge field is regular at the origin and behaves around the origin
\begin{equation}
  A_{0} (r) = V_{0}(r) = \widetilde{a} (r)  \approx g v A_{1} r  \ \ \ \ (r \approx 0)
,
\end{equation}
although the time component $\mathscr{X}_{0} (x)$ of the massive mode is absent, $\mathscr{X}_{0} (x) = 0$ and the time component $A_{0} (r)$ is identical to the restricted field $V_{0} (r)$.

By choosing the gauge transformation matrix $ U(x)$ as
\begin{align}
U (x) = \begin{pmatrix}
\cos \frac{\theta}{2} & \sin \frac{\theta}{2} e^{- i \varphi} \\
- \sin \frac{\theta}{2} e^{i \varphi} & \cos \frac{\theta}{2} 
\end{pmatrix} \in SU (2)
,
\end{align}
the fields are transformed into
\begin{align}
g \mathscr{V}_{0}^{\prime} (x) = & \widetilde{a} (r) T_{3} , \ \ \ 
g \mathscr{V}_{r}^{\prime} (x) = 0 , \ \ \ 
g \mathscr{V}_{\theta}^{\prime} (x) = 0 , \ \ \ 
g \mathscr{V}_{\varphi}^{\prime} (x) =  
\frac{1 - \cos \theta}{r\sin \theta} T_{3} , \\
g \mathscr{X}_{0}^{\prime} (x) = & 0 , \ \ \ 
g \mathscr{X}_{r}^{\prime} (x) = 0 , \ \ \ 
g \mathscr{X}_{\theta}^{\prime} (x) = - \frac{f (r)}{r} T_{+}, \ \ \ 
g \mathscr{X}_{\varphi}^{\prime} (x) = - \frac{f (r)}{r} T_{-}
,
\end{align}
where we have defined
\begin{align}
T_{+} := \frac{1}{2} \begin{pmatrix}
0 & - i e^{- i \varphi} \\
i e^{i \varphi} & 0 
\end{pmatrix} , \ \ \ 
T_{-} := \frac{1}{2} \begin{pmatrix}
0 & - e^{- i \varphi} \\
- e^{i \varphi} & 0 
\end{pmatrix}
.
\end{align}
We find that the dyonic contribution appears in the Wu--Yang potential $\mathscr{V} (x)$, while there are no effects in the massive mode $\mathscr{X} (x)$.

\section{Chromoelectric and chromomagnetic fields of a dyon}

In this section, all the expressions are given for the Julia--Zee dyon, i.e., radially variable case.
The expressions for the Yang--Mills (radially fixed) dyon can be easily obtained by setting $h (\rho) = 1$.

In the similar way to the Yang--Mills monopole \cite{Nishino}, we examine the magnetic charge $q_{m}$ and electric charge $q_{e}$ obtained by the chromomagnetic field $\mathscr{B}_{j}^{A} (x)$ and chromoelectric field $\mathscr{E}_{j}^{A} (x)$:
\begin{align}
g \mathscr{B}_{j}^{A} (x) := & \frac{1}{2} g \epsilon_{j k l} \mathscr{F}_{k l}^{A} (x) = 
\frac{x^{A} x^{j}}{r^{4}} \left( 1 - f^{2} ( r ) \right) - \left( \frac{\delta^{A j}}{r} - \frac{x^{A} x^{j}}{r^{3}} \right) \frac{d}{d r} f ( r ) , \\
g \mathscr{E}_{j}^{A} (x) := & g \mathscr{F}_{j 0}^{A} (x) = 
\frac{x^{A} x^{j}}{r^{2}} \frac{d}{d r} \widetilde{a} (r) + \left( \frac{\delta^{A j}}{r} - \frac{x^{A} x^{j}}{r^{3}} \right) \widetilde{a} (r) f ( r)
.
\end{align}
The magnetic charge $q_{m}$ and its density $\rho_{m} (r)$ are defined by
\begin{equation}
q_{m} = \int d^{3} x \ \mathscr{B}_{j}^{A} \left( \mathscr{D}_{j} [\mathscr{A}] \hat{\bm{\phi}} \right)^{A} = \frac{4 \pi}{g} \int_{0}^{\infty} d r \ \rho_{m} (r) , \ \ \ 
\rho_{m} (r) := g r^{2} \mathscr{B}_{j}^{A} \left( \mathscr{D}_{j} [ \mathscr{A} ] \hat{\bm{\phi}} \right)^{A} 
.
\end{equation}
Similarly, the electric charge $q_{e}$ and its density $\rho_{e} (r)$ are defined by
\begin{equation}
q_{e} = \int d^{3} x \ \mathscr{E}_{j}^{A} \left( \mathscr{D}_{j} [\mathscr{A}] \hat{\bm{\phi}} \right)^{A} = \frac{4 \pi}{g} \int_{0}^{\infty} d r \ \rho_{e} (r) , \ \ \ 
\rho_{e} (r) := g r^{2} \mathscr{E}_{j}^{A} \left( \mathscr{D}_{j} [\mathscr{A}] \hat{\bm{\phi}} \right)^{A}
.
\end{equation}
The charge densities $\rho_{m} (r)$ and $\rho_{e} (r)$ can be written in terms of the profile functions:
\begin{align}
\rho_{m} (r) = & \frac{d}{d r} \bigl[ h (r) \left( 1 - f^{2} (r) \right) \bigr] , \\
\rho_{e} (r) = & r^{2} \frac{d}{d r} \widetilde{a} (r) \frac{d}{d r} h (r) + 2 \widetilde{a} (r) h (r) f^{2} (r)
.
\end{align}

\begin{figure}[t]
\centering
\includegraphics[width=0.45\textwidth]{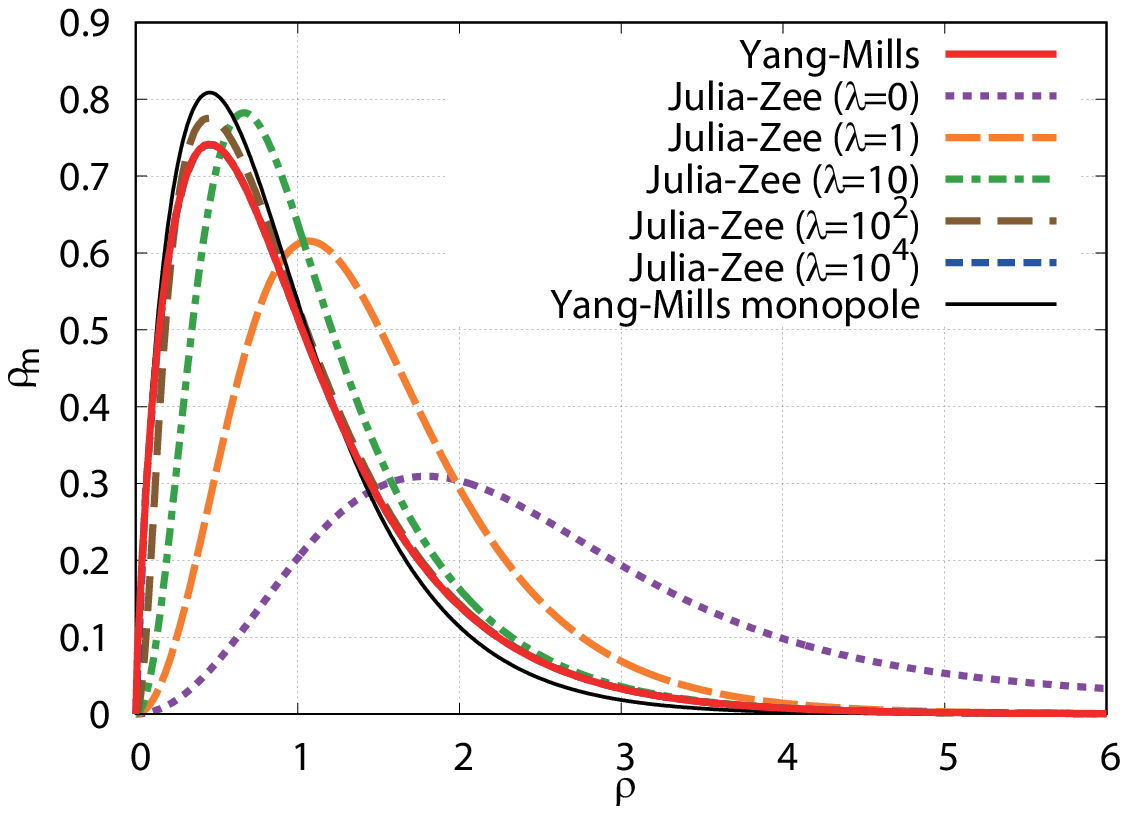} \ 
\includegraphics[width=0.45\textwidth]{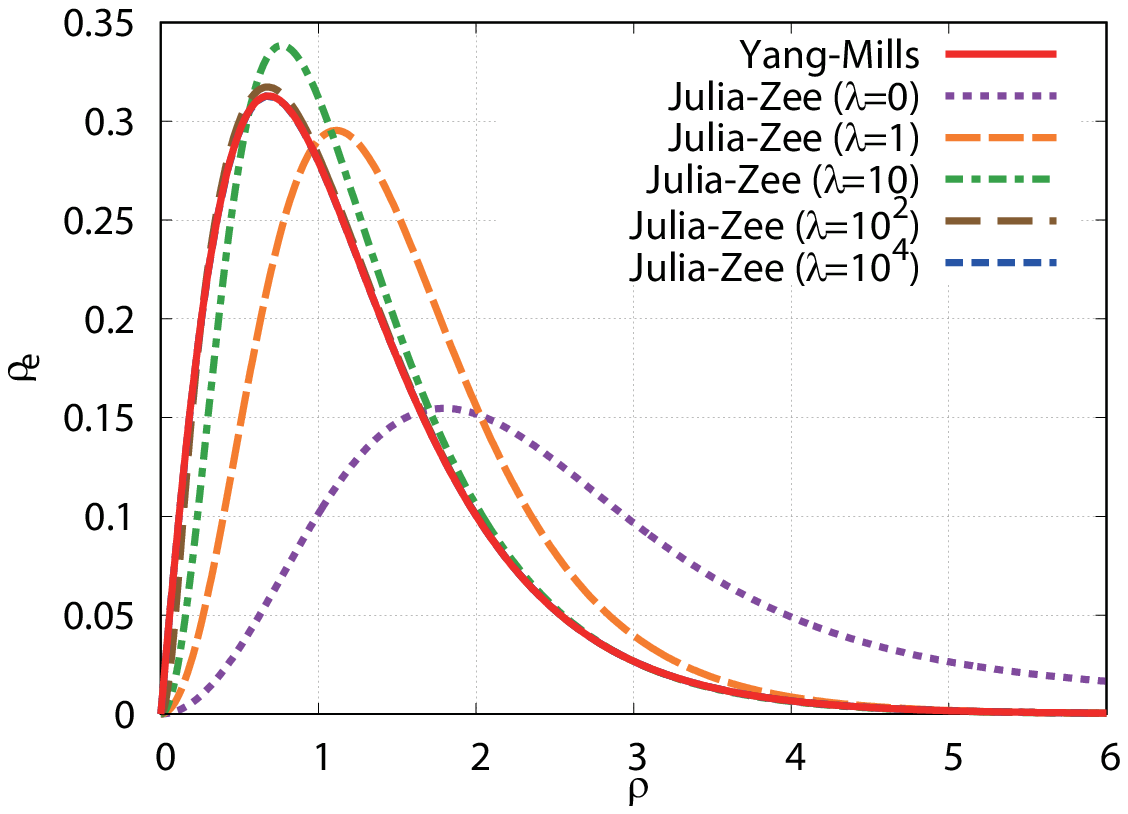}
\caption{(Left panel) The magnetic charge density $\rho_{m}$ and (Right panel) electric charge density $\rho_{e}$ as functions of $\rho = g v r$ at the $C = 0.5$.}
\label{charge_density}
\end{figure}

Fig. \ref{charge_density} is the plots of the charge densities $\rho_{m}$ and $\rho_{e}$ as functions of $\rho = g v r$ at  $C = 0.5$.

\begin{figure}[t]
\centering
\includegraphics[width=0.45\textwidth]{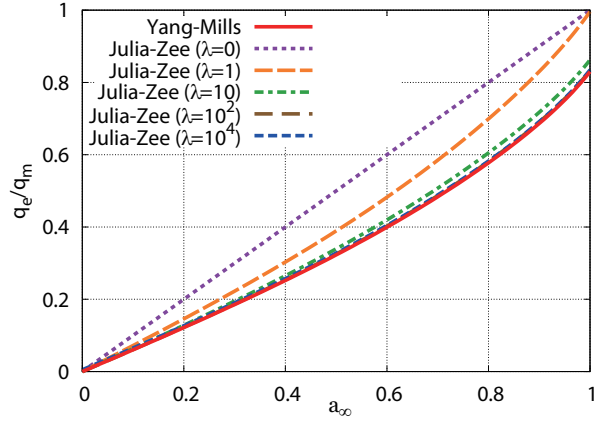} 
\caption{
The $a_{\infty}$ dependence of the ratio $q_{e} / q_{m}$ between two charges $q_{e},q_{m}$, namely $C$.
Throughout in this paper, we fix the magnetic charge $q_{m}$ to the unit $4 \pi / g$, which means that $C$ is nothing but the electric charge $q_{e}$ in unit of $4\pi/g$.}
\label{C_omega}
\end{figure}

We also illustrate the $a_{\infty}$-dependence of the ratio $C$ of the charges, $C = q_{e}/ q_{m}$, which is shown in Fig. \ref{C_omega}.

Next, we investigate the behavior of the chromomagnetic field $\mathscr{B}_{j}^{A} (x)$ and chromoelectric field $\mathscr{E}_{j}^{A} (x)$, especially around $r \approx 0$.
To do so, we return to the polar coordinate representation:
\begin{align}
g \mathscr{B}_{r} (x) = & \frac{1 - f^{2} (r)}{r^{2}}  T_{r}
,  \ \ \ 
g \mathscr{B}_{\theta} (x) =  \frac{1}{r} \frac{d f (r)}{d r} T_{\varphi}
, \ \ \ 
g \mathscr{B}_{\varphi} (x) =  -\frac{1}{r} \frac{d f (r)}{d r}  T_{\theta} \label{magnetic_field}
,\\
g \mathscr{E}_{r} (x) = & \frac{d \widetilde{a} (r)}{d r}  T_{r} , \ \ \ 
g \mathscr{E}_{\theta} (x) =  -\frac{\widetilde{a} (r) f (r)}{r} T_{\varphi} , \ \ \ 
g \mathscr{E}_{\varphi} (x) = \frac{\widetilde{a} (r) f (r)}{r} T_{\theta}
.\label{electric_field}
\end{align}
In order to define the gauge-invariant field strength, we take the inner product between (\ref{magnetic_field}) or (\ref{electric_field}) and $\hat{\bm{\phi}} (x)$,
\begin{equation}
g \mathcal{B}_{r} (x) := g \mathscr{B}_{r} (x) \cdot \hat{\bm{\phi}} (x) = \frac{1 - f^{2} (r)}{r^{2}} , \ \ \ 
g \mathcal{E}_{r} (x) := g \mathscr{E}_{r} (x) \cdot \hat{\bm{\phi}} (x) = \frac{d}{d r} \widetilde{a} (r)
,
\end{equation}
and the other components are zero.

\begin{figure}[t]
\centering
\includegraphics[width=0.45\textwidth]{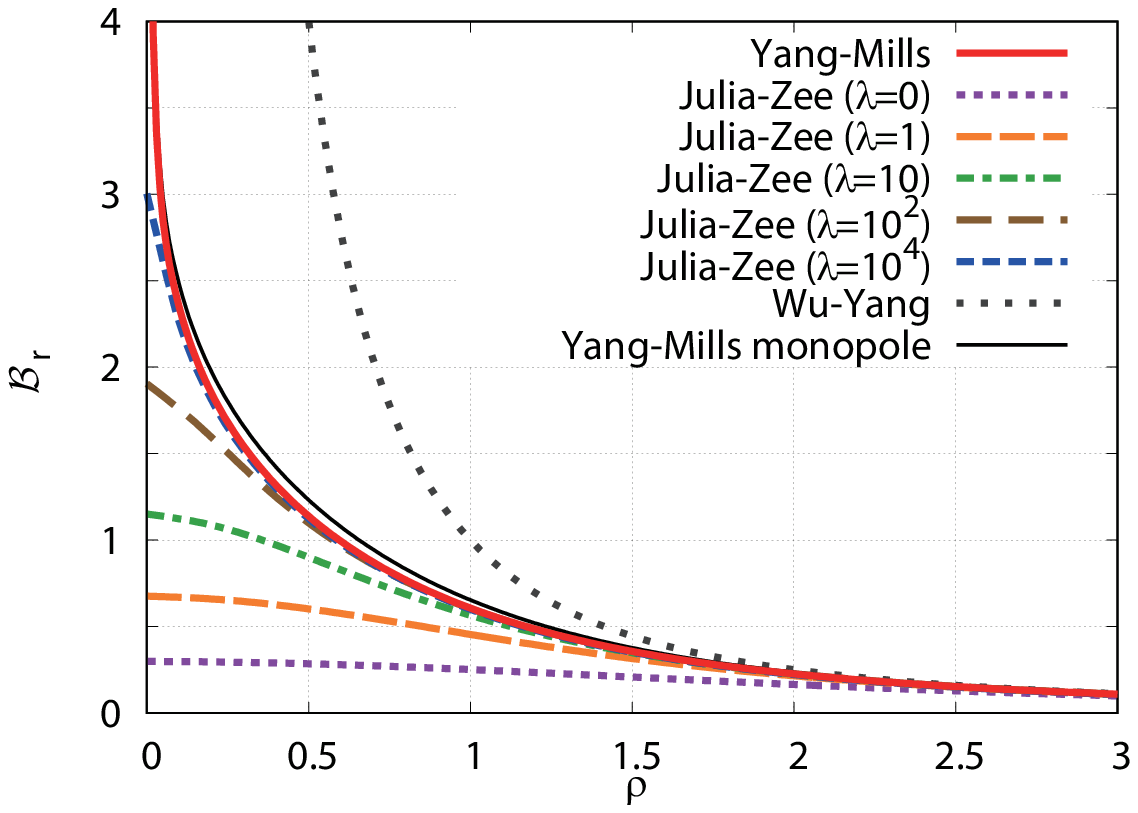} \ 
\includegraphics[width=0.45\textwidth]{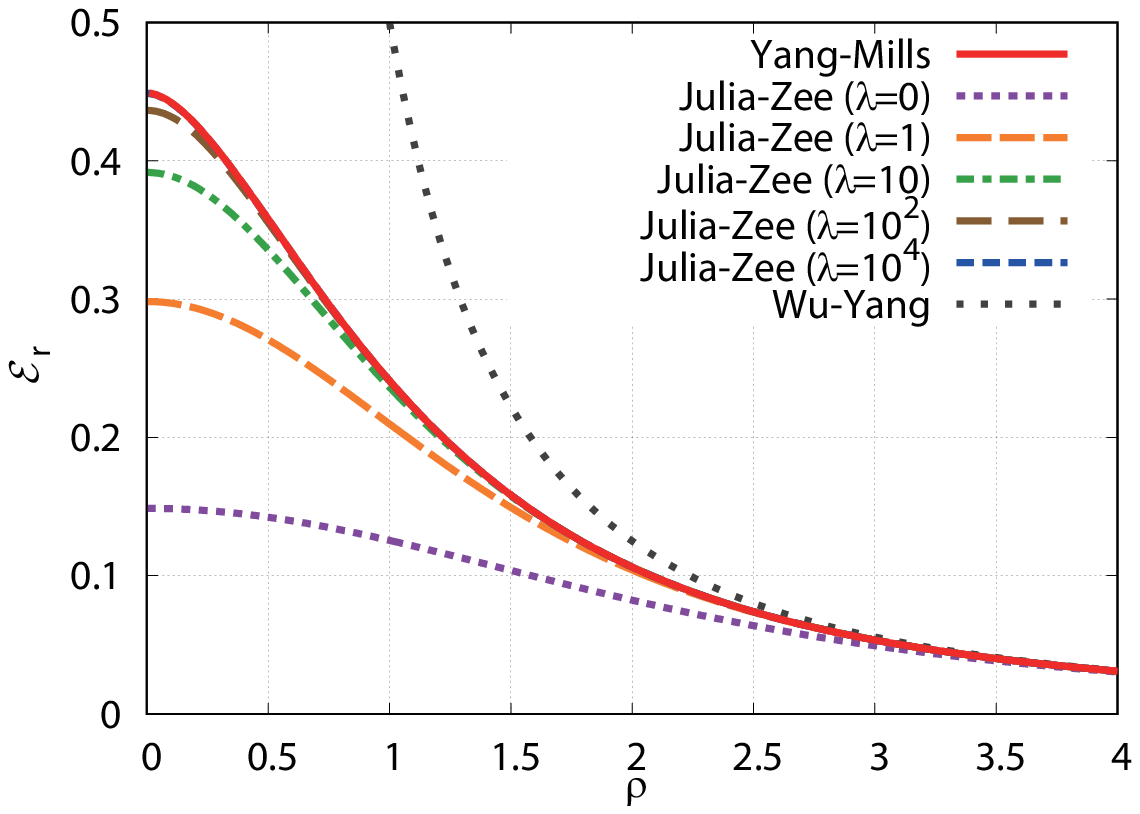}
\caption{The short-distance behaviors of (left panel) the gauge-invariant chromomagnetic field $\mathcal{B}_{r}$ and (right panel) the gauge-invariant chromoelectric field $\mathcal{E}_{r}$ as functions of $\rho = g v r$ at $C = 0.5$.
The chromomagnetic field of the Yang--Mills dyon diverges at the origin logarithmically, just like the Yang--Mills monopole.}
\label{dyon_B_E}
\end{figure}

Fig. \ref{dyon_B_E} is the plot of the gauge-invariant chromomagnetic and chromoelectric field strengths as functions of $\rho = g v r$ at $C= 0.5$.
The chromoelectric field $\mathcal{E}_{r} (x)$ is regular at the origin even for the Yang--Mills dyon.
However, the chromomagnetic field $\mathcal{B}_{r} (x)$ of the Yang--Mills dyon diverges logarithmically at the origin, just like the Yang--Mills monopole.

\section{Energy density and static mass of a dyon }

We define the energy integral $I$ as a function of $a_{\infty}$ and $\lambda$ by integrating the energy density $e (\rho)$ defined by (\ref{energy_density})
\begin{equation}
I (a_{\infty}  , \lambda ) = \int_{0}^{\infty} d \rho \ e (\rho)
,
\label{energy_integral}
\end{equation}
so that the energy $E$  takes the form:
\begin{equation}
E = \frac{4 \pi M_{\mathscr{X}}}{g^{2}} I (a_{\infty} , \lambda) 
.
\label{energy_integral2}
\end{equation}

\begin{figure}[t]
\centering
\includegraphics[width=0.45\textwidth]{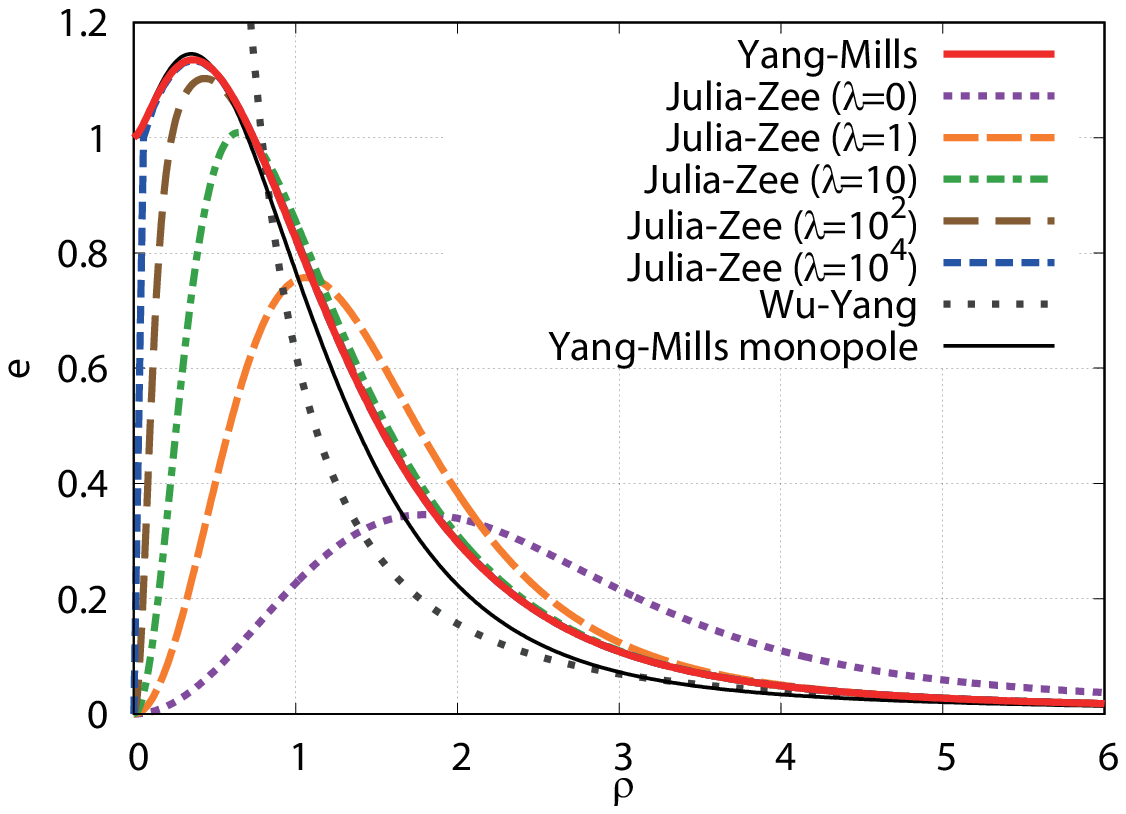} \ 
\includegraphics[width=0.45\textwidth]{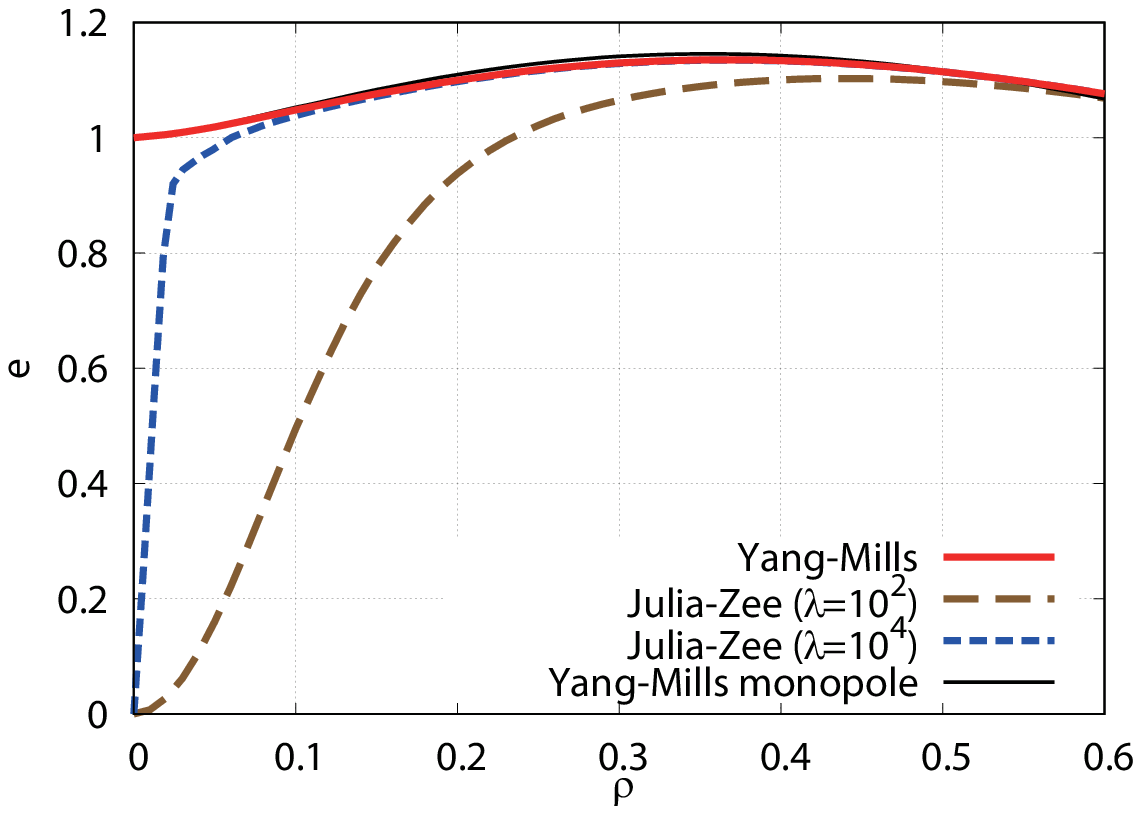}
\caption{The energy density $e$ of the Yang--Mills dyon as a function of $\rho = g v r$ at $C = 0.5$ to be compared with the Julia--Zee dyons for $\lambda = 0, 1, 10, 10^{2}$, and $10^{4}$, where the Wu--Yang monopole and the Yang--Mills monopole are also plotted for reference.
(Left panel) $0 \leq \rho \leq 6$, (Right panel) $0 \leq \rho \leq 0.6$.}
\label{dyon_energy_1}
\end{figure}

Fig. \ref{dyon_energy_1} is the plot of the energy density $e (\rho)$ as a function of $\rho$ obtained from the solution $a (\rho), f (\rho)$ and $h(\rho)$ at $C = 0.5$, which should be compared with the Julia--Zee solution.
We find that at the origin the energy density $e (\rho)$ of the Yang--Mills dyon takes the value $e (0) = 1$, while the Julia--Zee dyons behave $e (0) = 0$ for any values of $0 \leq \lambda < \infty$.
This difference is caused by the radially fixing condition $h (\rho) = 1$.
The sixth term $h^{2} (\rho) f^{2} (\rho)$ in (\ref{energy_density}) survives at the origin in the Yang--Mills dyon due to $h (0) = 1$, while for the Julia--Zee dyon it vanishes since $h (0) = 0$.

In the BPS limit $\lambda = 0$ of the 't Hooft--Polyakov monopole ($a_{\infty} = 0$), the integral $I$ takes the value one:
\begin{equation}
I (a_{\infty} = 0 , \lambda = 0) = 1
,
\label{energy_integral_I}
\end{equation}
which leads to 
\begin{equation}
E = \frac{4 \pi M_{\mathscr{X}}}{g^{2}}
.
\end{equation}

\begin{figure}[t]
\centering
\includegraphics[width=0.45\textwidth]{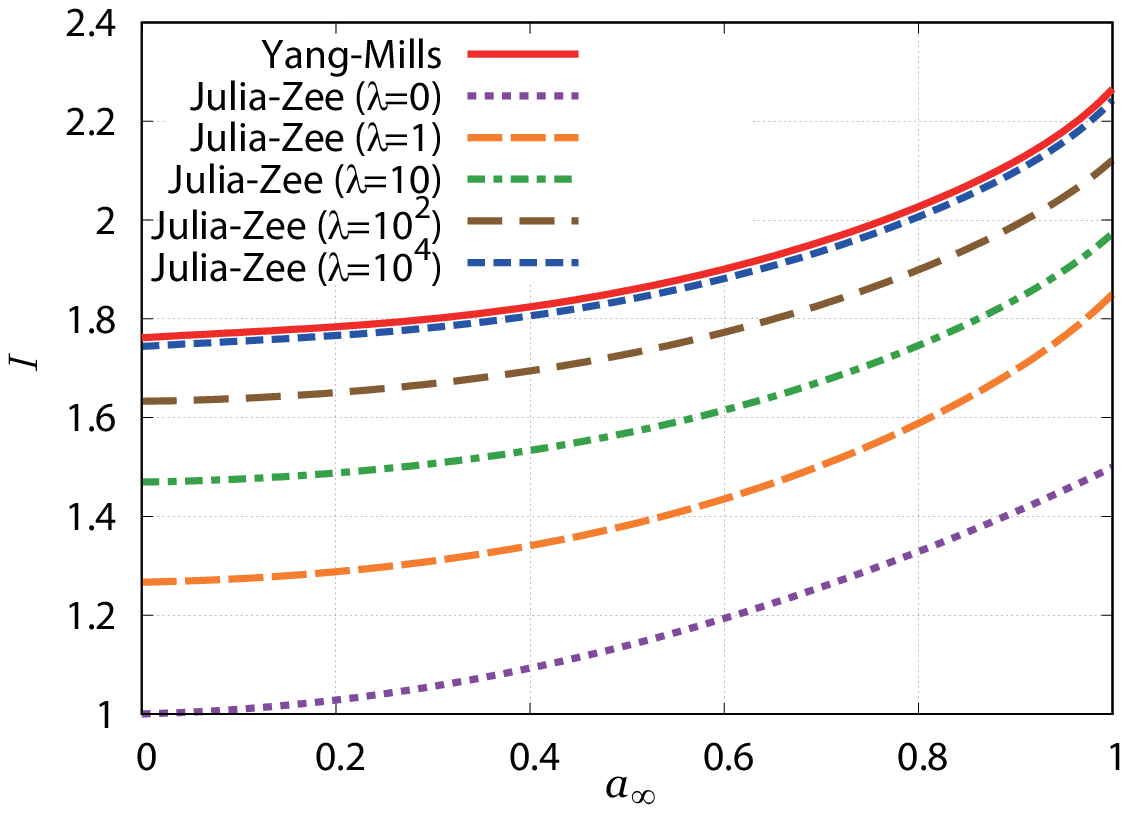} \ 
\includegraphics[width=0.45\textwidth]{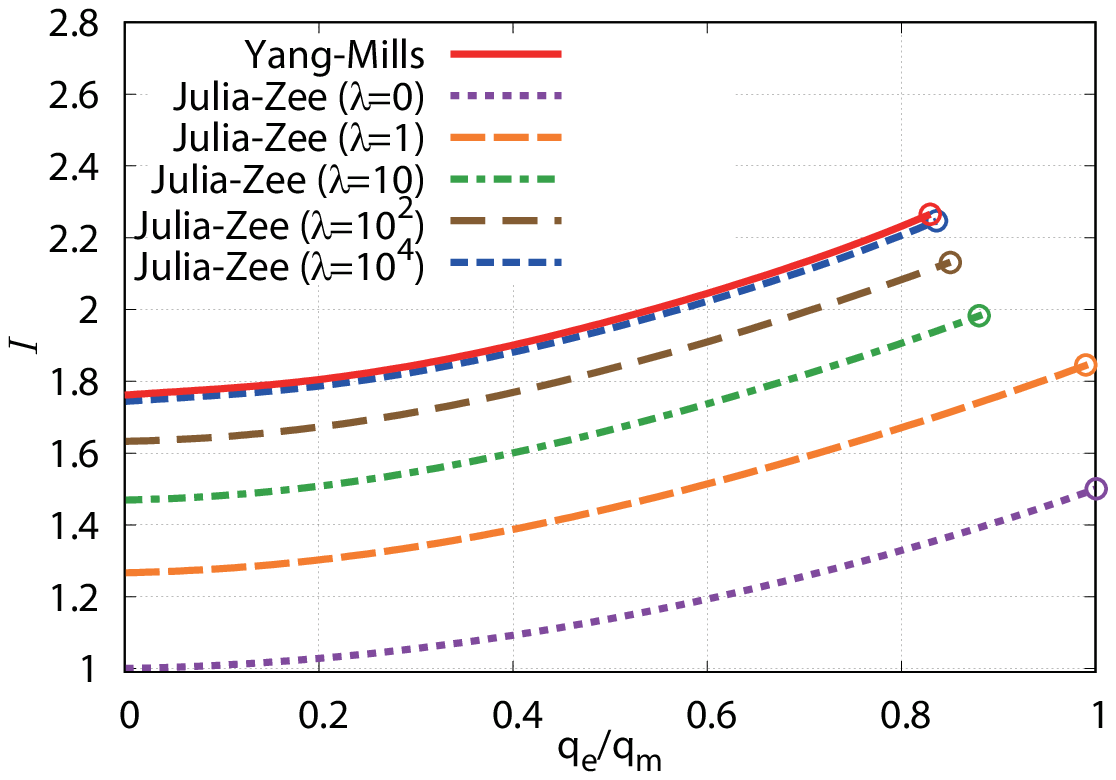}
\caption{(Left panel) The $a_{\infty}$ dependence of the energy integral $I$ for various values of $\lambda$.
(Right panel) The plot of the energy integral $I$ versus the ratio of the charges $q_{e} / q_{m}$.
The circles stand for end points of the existence of the dyon, namely $a_{\infty} = 1$.}
\label{dyon_energy_2}
\end{figure}

For a fixed value of the scalar coupling $\lambda$, the energy $E$ is monotonically increasing in $a_{\infty}$ as seen in the left panel of Fig. \ref{dyon_energy_2} or in $C = q_{e}/ q_{m}$ as seen in the right panel of Fig. \ref{dyon_energy_2}.
The right panel of Fig. \ref{dyon_energy_2} has been obtained in \cite{Bais-Primack} for the Julia--Zee dyon with finite values of $\lambda$ and we have added the Yang--Mills dyon, which partially corresponds to the limit of $\lambda \to \infty$ of the Julia--Zee dyon.
For the Yang--Mills dyon and the Julia--Zee dyon for a sufficiently large coupling $\lambda \gg 1$ above a critical value $\lambda^{\prime}$ of $\lambda$, the electric charge $q_{e}$ cannot reach the maximal limit $q_{e}/q_{m} = 1$: the maximal value of $q_{e}/q_{m}$ is obtained in a numerical way as
\begin{equation}
\frac{q_{e}}{q_{m}} \biggl|_{a_{\infty} \to 1, \lambda \to \infty} = 0.828
.
\end{equation}

By using the maximal value of the energy integral for the Yang--Mills dyon,
\begin{equation}
I ( a_{\infty} \to 1, \lambda \to \infty) = 2.265
,
\end{equation}
the maximal value of the static mass of the Yang--Mills dyon can be estimated as
\begin{equation}
E = \frac{4 \pi M_{\mathscr{X}}}{g^{2}} I (a_{\infty} \to 1, \lambda \to \infty) = 1.18 \pm 0.05 \ \mathrm{GeV}
,
\end{equation}
where we have used the value for the off-diagonal gluon mass $M_{\mathscr{X}} = 1.2 \mathrm{GeV}$ obtained by the preceding studies on a lattice \cite{Shibata2007}  and 
 the typical value of the running coupling constant $\alpha_{s} (p) := g^{2} (p)/4\pi \approx 2.3 \pm 0.1$ at $p \simeq M_{\mathscr{X}}\approx 1.2$GeV obtained in \cite{Bloch2003}.

The Yang--Mills dyon mass, $1.18 \mathrm{GeV}$, obtained in this paper is equal to the heaviest one in the family of Julia--Zee dyons in the Georgi--Glashow model, since the energy integral (\ref{energy_integral}) is monotonically increasing in the coupling constant $\lambda$ and the asymptotic value $a_{\infty}$ of the time component of the gauge field.
The Yang--Mills dyon mass, $1.18 \mathrm{GeV}$, is $27 \%$ larger than the Yang--Mills monopole one: $0.93 \mathrm{GeV}$ found in \cite{Nishino}, it still remains in the same order of the off-diagonal gluon mass: $M_{\mathscr{X}} = 1.2 \mathrm{GeV}$.
In view of these, the existence of the Yang--Mills dyon with a reasonable mass tells us that the  Yang--Mills dyons can play the role of the quark confiner instead of the Yang--Mills monopoles if  their condensations occur according to the dual superconductor picture.

\section{Confinement/deconfinement phase transition and the Yang--Mills dyons}

In this section, we consider the Yang--Mills dyons on $S^{1} \times \mathbb{R}^{3}$ space.  
We introduce the coordinates $x=(\tau,\bm{x}) \in S^{1} \times \mathbb{R}^{3}$ with the ``time''  coordinate $\tau = x_{4}$ for $S^{1}$ and spatial coordinates $\bm{x}=(x_{1}, x_{2} , x_{3})$ for $\mathbb{R}^{3}$. 
Suppose we have performed the Wick rotation to the coordinates and the field  in the Minkowski spacetime to obtain the Euclidean counterparts: $\tau = i x_{0}$ and $\mathscr{A}_{4} (\tau , \bm{x})=-i\mathscr{A}_{0} (x_{0} , \bm{x})$.
Then the Euclidean action $S_{\rm E}^{\rm mYM}$ is obtained as
\begin{equation}
S_{\rm E}^{\rm mYM} = \int_{0}^{T^{-1}} d \tau \int d^{3} \bm{x} \  \biggl[ \frac{1}{4} \mathscr{F}_{\mu \nu} \cdot \mathscr{F}_{\mu \nu} + \frac{1}{2} \left( \mathscr{D}_{\mu} [\mathscr{A}] \bm{\phi} \right) \cdot \left( \mathscr{D}_{\mu} [\mathscr{A}] \bm{\phi} \right) + u \left( \bm{\phi} \cdot \bm{\phi} - v^{2} \right) \biggr]
,
\label{Euclid_action}
\end{equation}
where Greek indices $\mu$ and $\nu$ run from $1$ to $4$.
We have introduced the period $T^{-1}$ of the ``time'' direction $\tau$, which is regarded as the inverse of temperature. 
Note that the fields on $S^1 \times \mathbb{R}^3$ must be periodic in $\tau$ with the period $T^{-1}$ on $S^1$, that is to say, the fields must satisfy the periodic boundary condition 
\begin{equation}
 \mathscr{A}_{\mu}^{A} (\tau+T^{-1},\bm{x})=\mathscr{A}_{\mu}^{A} (\tau,\bm{x}), \quad
\hat{\phi}^{A} (\tau+T^{-1},\bm{x})=\hat{\phi}^{A} (\tau,\bm{x}).
\label{p-bc}
\end{equation}

\subsection{The self-dual dyon and the KvBLL caloron in the massless Yang--Mills theory}

Before obtaining the Yang--Mills dyon solution, let us review the conventional dyon solution in the pure massless $SU(2)$ Yang--Mills theory on $S^{1} \times \mathbb{R}^{3}$ space.
The action $S_{\rm E}^{\rm YM}$ is given by removing the scalar field $\bm{\phi}$ from (\ref{Euclid_action}):
\begin{equation}
S_{ \rm E}^{\rm YM} = \int_{0}^{T^{-1}} d \tau \int d^{3} \bm{x} \ \frac{1}{4} \mathscr{F}_{\mu \nu} \cdot \mathscr{F}_{\mu \nu}
.
\label{S_E_YM}
\end{equation}
This action is nonnegative and has a lower bound:
\begin{align}
S_{\rm E}^{\rm YM} = & \int_{0}^{T^{-1}} d \tau  \int d^{3} \bm{x} \ \biggl[ \frac{1}{8} \left( \mathscr{F}_{\mu \nu} \mp {}^{\star} \mathscr{F}_{\mu \nu} \right)^{2} \pm \frac{1}{4} \mathscr{F}_{\mu \nu} \cdot {}^{\star} \mathscr{F}_{\mu \nu} \biggr] \nonumber\\
\geq & \biggl|  \pm \int_{0}^{T^{-1}} d \tau \int d^{3} \bm{x} \ \frac{1}{4} \mathscr{F}_{\mu \nu} \cdot {}^{\star} \mathscr{F}_{\mu \nu} \biggr|
,
\label{lower_bound_action}
\end{align}
where we have introduced the Hodge dual ${}^{\star} \mathscr{F}_{\mu \nu}$ of the field strength tensor $\mathscr{F}_{\mu \nu}$ by
\begin{equation}
{}^{\star} \mathscr{F}_{\mu \nu} := \frac{1}{2} \epsilon_{\mu \nu \alpha \beta} \mathscr{F}_{\alpha \beta}
.
\end{equation} 
The equality of (\ref{lower_bound_action}) holds if and only if the equation
\begin{equation}
\mathscr{F}_{\mu \nu} = \pm {}^{\star} \mathscr{F}_{\mu \nu}
,
\label{self_dual_original}
\end{equation}
is satisfied.
The equation (\ref{self_dual_original}) is called the {\it self-dual equation} for the plus sign on the right hand side, and the {\it anti-self-dual equation} for the minus sign.
In what follows, we will concentrate the self-dual equation.

First, we adopt the ansatz for $S^{1} \times \mathbb{R}^{3}$ space 
\begin{equation}
g \mathscr{A}_{j}^{A} (\tau , \bm{x}) = \epsilon^{j A k} \frac{x^{k}}{r} \frac{1 - \widetilde{f} (r)}{r} , \ \ \ 
g \mathscr{A}_{4}^{A} (\tau,\bm{x}) = \frac{x^{A}}{r} \widetilde{a} (r)
,
\label{E_ansatz_sd}
\end{equation}
where Roman indices $j$ and $k$ run from $1$ to $3$ and $r$ is the radius in $\mathbb{R}^{3}$, i.e., $r = \sqrt{x_{j} x_{j}}$. 
In fact, the ansatz (\ref{E-ansatz}) is $\tau$-independent and hence it trivially satisfies the periodic boundary condition (\ref{p-bc}). 

The nontrivial components of the self-dual equation (\ref{self_dual_original}) with the ansatz (\ref{E_ansatz_sd}) is given by
\begin{equation}
\mathscr{F}_{j 4}^{A} = \frac{1}{2} \epsilon_{j k l} \mathscr{F}_{k l}^{A}
,
\label{self-dual-eq}
\end{equation}
which is written in terms of the profile functions $\widetilde{a} (r)$ and $\widetilde{f} (r)$ as
\begin{equation}
r^{2} \widetilde{a}^{\prime} (r) = 1 - \widetilde{f}^{2} (r) , \ \ \ \widetilde{f}^{\prime} (r) = - \widetilde{a} (r) \widetilde{f} (r)
\label{self-dual-a-f}
.
\end{equation}

The solution of the self-dual equations (\ref{self-dual-a-f}) is exactly obtained as
\begin{align}
\widetilde{a} (r) = & V \left( \coth (V r) - \frac{1}{V r} \right)  \xrightarrow{r \to \infty} V - \frac{1}{r} 
, \label{sd_sol_a2}\\ 
\widetilde{f} (r) = & \frac{V r}{\sinh \left( V r \right)} \xrightarrow{r \to \infty} 0
,
\label{sd_sol_f2}
\end{align}
where $V > 0$ is an arbitrary parameter with a dimension of mass, which is related to the asymptotic holonomy.
For later convenience, we set 
\begin{equation}
V = g v a_{\infty} = \widetilde{a}_{\infty}
\label{parameter_relation}
.
\end{equation}

By introducing the dimensionless variable $\rho$ and functions $a$ and $f$ in the same way as (\ref{dimensionless}) and using the relation (\ref{parameter_relation}), the solutions (\ref{sd_sol_a2}) and (\ref{sd_sol_f2}) are cast into
\begin{align}
a (\rho) = & a_{\infty} \left( \coth ( a_{\infty} \rho ) - \frac{1}{a_{\infty} \rho} \right) \xrightarrow{\rho \to \infty} a_{\infty} - \frac{1}{\rho} 
, \label{sd_sol_a}\\
f (\rho) = & \frac{a_{\infty} \rho}{\sinh \left( a_{\infty} \rho \right)} \xrightarrow{\rho \to \infty} 0
.
\label{sd_sol_f}
\end{align}
Notice that the solutions (\ref{sd_sol_a}) and (\ref{sd_sol_f}) of the self-dual equations (\ref{self-dual-a-f}) also satisfy the second-order field equations
\begin{align}
& a^{\prime \prime} (\rho) + \frac{2}{\rho} a^{\prime} (\rho) - \frac{2}{\rho^{2}} a (\rho) f^{2} (\rho) = 0 , \label{sd_eq_a}\\
& f^{\prime \prime} (\rho) - \frac{1}{\rho^{2}} \left( f^{3} (\rho) - f (\rho) \right) - a^{2} (\rho) f (\rho) = 0
,
\label{sd_eq_f}
\end{align}
which is obtained by substituting the ansatz (\ref{E_ansatz_sd}) to the Yang--Mills field equation as suggested from (\ref{GG_eq_a_dyon}) and (\ref{GG_eq_f_dyon}) by the replacement $a \to - i a$ and $h \to 0$.

By recalling the asymptotic behavior (\ref{YM_dyon_asymptotic_form}) of the profile function $ a(\rho)$, the ratio $C$ of the charges for the self-dual dyon is fixed
\begin{equation}
C = \frac{q_{e}}{q_{m}} \equiv 1
,
\label{C_sd}
\end{equation}
for any values of $a_{\infty}$.

The action $S_{\rm E}^{\rm YM}$ is also rewritten as
\begin{equation}
S_{\rm E}^{\rm YM} = \frac{4 \pi}{g^{2}} \frac{g v}{T} I (a_{\infty})
,
\end{equation}
where we have defined the dimensionless energy $I$ by
\begin{equation}
I (a_{\infty}) := \int_{0}^{\infty} d \rho \ \biggl[ \frac{1}{2} \rho^{2} a^{\prime 2} (\rho) + a^{2} (\rho) f^{2} (\rho) + f^{\prime 2} (\rho) + \frac{\left( f^{2} (\rho)  -1 \right)^{2}}{2 \rho^{2}} \biggr]
.
\label{energy_integral_sd}
\end{equation}
The dimensionless energy $I$ is calculated by using the solutions (\ref{sd_sol_a}) and (\ref{sd_sol_f}) as
\begin{equation}
I (a_{\infty}) = a_{\infty}
,
\end{equation}
and hence the action $S_{\rm E}^{\rm YM}$ reads
\begin{equation}
S_{\rm E}^{\rm YM} = \frac{4 \pi}{g^{2}} \frac{g v a_{\infty}}{T} = \frac{4 \pi}{g^{2}} \frac{V}{T}
.
\label{action_value_sd}
\end{equation}

The gauge field $\mathscr{A}_{\mu} (x)$ for the self-dual dyon is asymptotically written as
\begin{align}
g \mathscr{A}_{r} \approx 0, \ \ \  g \mathscr{A}_{\theta} \approx 0 , \ \ \ 
g \mathscr{A}_{\varphi} \approx  \frac{1 - \cos \theta}{r \sin \theta} T_{3}   , \ \ \ 
g \mathscr{A}_{4} \approx  \left( V - \frac{1}{r} \right) T_{3}
, \ \ \ \left(r \gg V^{-1} \right) ,
\label{asymptotic_sd}
\end{align}
by performing the gauge transformation to the unitary gauge (or the stringy gauge) in which the $\mathscr{A}_{4}$ component is constant and diagonal at spatial infinity.

\begin{figure}[t]
\centering
\includegraphics[width=0.6\textwidth]{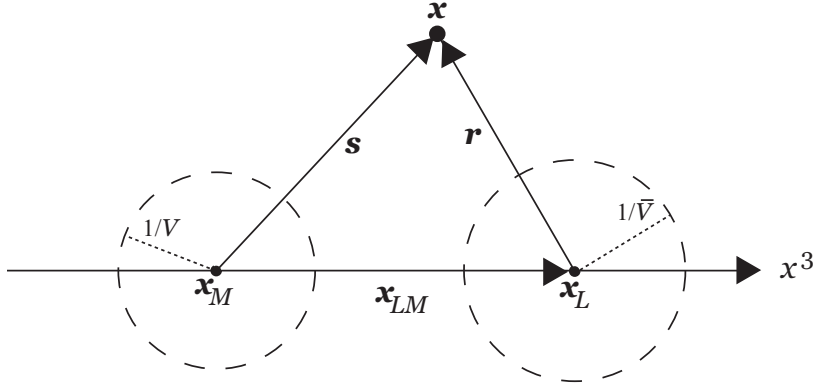}
\caption{The coordinates of a KvBLL caloron in terms of the two constituent dyons $L,M$.}
\label{KvBLL_fig}
\end{figure}

Second, the pure massless $SU(2)$ Yang--Mills theory has an another topological soliton solution.
The {\it KvBLL calorons} are the solution of the (anti-)self-dual equation (\ref{self_dual_original}) of the pure massless Yang--Mills theory in $S^{1} \times \mathbb{R}^{3}$ space with a nontrivial holonomy.
Indeed, the gauge field $\mathscr{A}_{\mu} (x)$ of a KvBLL caloron is given by \cite{KvBLL}
\begin{equation}
g \mathscr{A}_{\mu} (\tau,\bm{x}) =  V \delta_{\mu 4} T_{3} + \bar{\eta}^{3}_{\mu \nu} \partial_{\nu} \log \frac{\psi}{\hat{\psi}} T_{3} 
+ \frac{\psi}{\hat{\psi}} \mathrm{Re} \bigl[ \left( \bar{\eta}^{1}_{\mu \nu} - i \bar{\eta}^{2}_{\mu \nu} \right) \left( T_{1} + i T_{2} \right) \left( \partial_{\nu} + i V \delta_{\nu 4} \right) \widetilde{\chi} \bigr]
,
\label{KvBLL}
\end{equation}
with three functions $\hat{\psi}$, $\psi$, and $\widetilde{\chi}$ defined by
\begin{align}
\hat{\psi} = & - \cos \left( 2 \pi T \tau \right) + \cosh \left( \bar{V} r \right) \cosh \left( V s \right) + \frac{r^{2} + s^{2} - \pi^{2} \varrho^{4} T^{2}}{2 r s} \sinh \left( \bar{V} r \right) \sinh \left( V s \right) , \\
\psi = & - \cos \left( 2 \pi T \tau \right) + \cosh \left( \bar{V} r \right) \cosh \left( V s \right) + \frac{r^{2} + s^{2} + \pi^{2} \varrho^{4} T^{2}}{2 r s} \sinh \left( \bar{V} r \right) \sinh \left( V s \right) \nonumber\\
& + \pi \varrho^{2} T \biggl[ \frac{\sinh \left( V s \right) \cosh \left( \bar{V} r \right)}{s} + \frac{\sinh \left( \bar{V} r \right) \cosh \left(  V s \right)}{r} \biggr] , \\
\widetilde{\chi} = & \frac{\pi \varrho^{2} T}{\psi} \biggl[ e^{- 2 \pi i T \tau} \frac{\sinh \left( V s \right)}{s} + \frac{\sinh \left( \bar{V} r \right)}{r} \biggr]
,
\end{align}
where $\bar{\eta}^{A}_{\mu \nu}$ is the 't Hooft symbol defined by $\bar{\eta}^{A}_{\mu \nu}=\epsilon_{A\mu\nu}-\delta_{A\mu}\delta_{\nu 4}+\delta_{A\nu}\delta_{\mu 4}$ 
and $\bar{V}:=2\pi T-V$. 
This solution indeed has the periodicity $T^{-1}$, i.e., $\mathscr{A}_{\mu} (\tau+T^{-1},\bm{x})=\mathscr{A}_{\mu} (\tau,\bm{x})$. 
 Let $\bm{x}_{M}$ be the location of the $M$ dyon's center of mass  and $\bm{x}_{L}$ the location of the $L$ dyon's center of mass in $\mathbb{R}^3$. 
For the observation point $\bm{x}$, we introduce the vector $\bm{s}:=\bm{x} - \bm{x}_{M}$ and  $\bm{r}:=\bm{x} - \bm{x}_{L}$ from the locations of the $M$ and $L$ dyons to the point $\bm{x}$ respectively.
Therefore, $s = |\bm{s}|= | \bm{x} - \bm{x}_{M}|$ and $r = |\bm{r}|= | \bm{x} - \bm{x}_{L}|$ are  respectively the distances from $M$ and $L$ dyons located at $\bm{x}_{M}$ and $\bm{x}_{L}$ to the observation point $\bm{x}$. 
The relative position vector $\bm{x}_{L M} := \bm{x}_{L}-\bm{x}_{M}$ between the two dyons is given by $\bm{x}_{L M} = \pi \varrho^{2} T \bm{e}_{3}$ with $\varrho$ being the size of the Belavin--Polyakov--Schwartz--Tyupkin (BPST) instanton \cite{BPST}  
where the direction is chosen to be along the third spatial direction. 
The core sizes of the two dyons $M,L$ are respectively given by $V^{-1}$ and $\bar{V}^{-1}$.
See Fig.~\ref{KvBLL_fig}.

The nontrivial holonomy of the KvBLL caloron  originates from the first term of (\ref{KvBLL}) for $V \not=0$.
If we take $V=0$, the KvBLL caloron reduces to the Harrington--Shepard (HS) caloron solution with a trivial holonomy \cite{Harrington-Shepard}.
In the zero temperature limit $T \to 0$, the KvBLL caloron reduces to the BPST instanton \cite{BPST} of size $\varrho$.

The constituent dyon can be identified if it is in the vicinity of one of its constituent dyons and far away from the other, namely, at large separations. 
For instance, near the $M$ dyon center and far away from the $L$ dyon $(r \gg 1/\bar{V})$, the KvBLL caloron solution (\ref{KvBLL}) exhibits the asymptotic behavior of the $M$ dyon
\begin{align}
g \mathscr{A}_{r} \approx 0, \ \ \  g \mathscr{A}_{\theta} \approx 0 , \ \ \ 
g \mathscr{A}_{\varphi} \approx  \frac{1 - \cos \theta}{s \sin \theta} T_{3}   , \ \ \ 
g \mathscr{A}_{4} \approx  \left( V - \frac{1}{s} \right) T_{3}
, \ \ \ \left(s \gg V^{-1} \right) ,
\label{asymptotic}
\end{align}
by performing the gauge transformation to the unitary gauge in which the $\mathscr{A}_{4}$ component is constant and diagonal at spatial infinity.
This is nothing but the self-dual dyon (\ref{asymptotic_sd}).

\subsection{The Yang--Mills dyon on $S^{1} \times \mathbb{R}^{3}$ space}

\begin{figure}[t]
\centering
\includegraphics[width=0.45\textwidth]{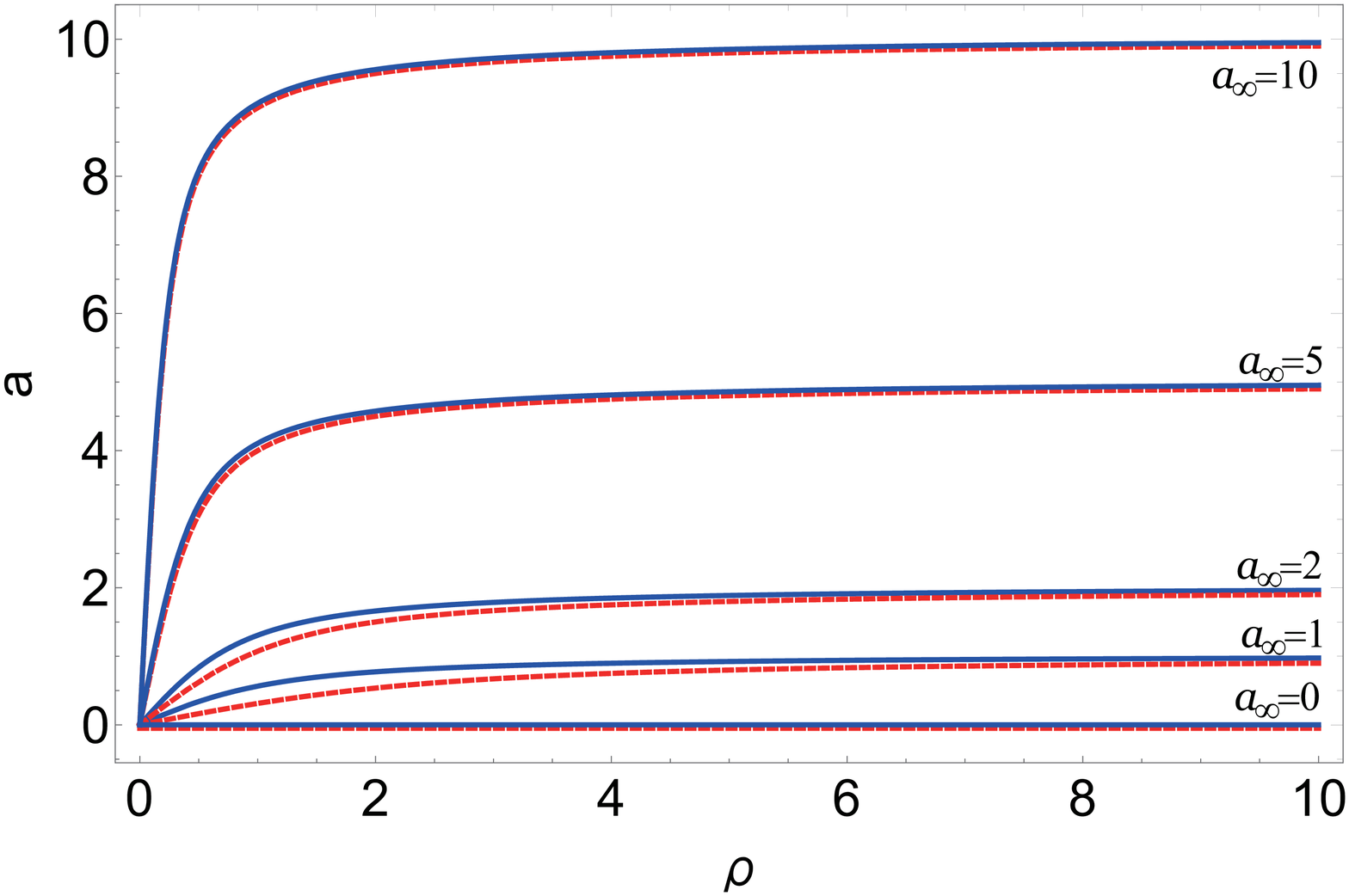} \ \ \
\includegraphics[width=0.45\textwidth]{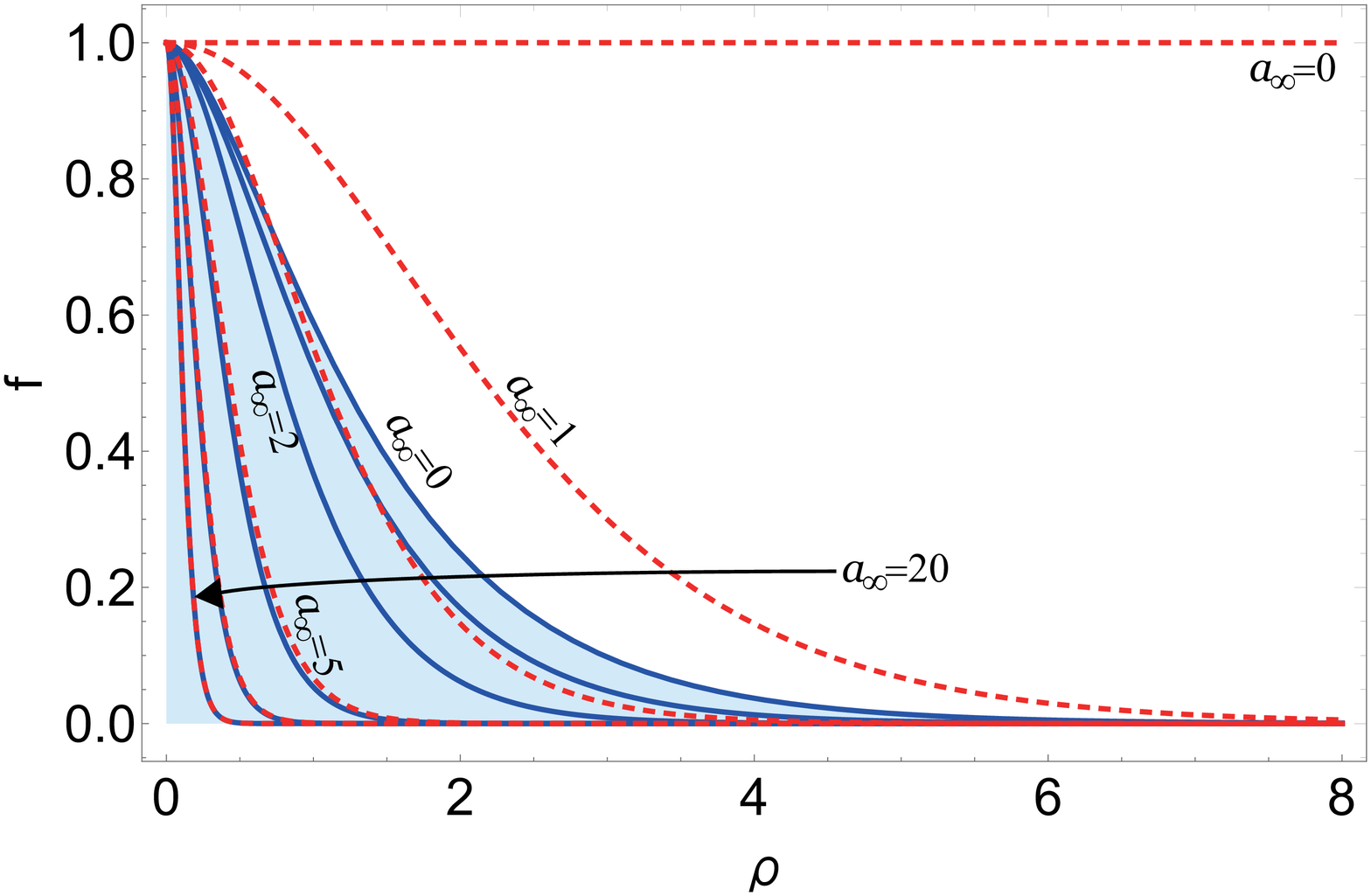}
\caption{The solutions $a$ (left panel) and $f$ (right panel) of the field equations (\ref{E_eq_a2}) and (\ref{E_eq_f2}) as functions of $\rho$.
The broken lines are the profile functions (\ref{sd_sol_a}) and (\ref{sd_sol_f}) of the self-dual dyon as a constituent of the KvBLL caloron.
The solid lines are the profile functions of the Yang--Mills dyon for various values of $a_{\infty}$ from $a_{\infty} = 0$ to $a_{\infty} \to \infty$ which covers the shaded region in the right panel.}
\label{YM_dyon_Euclid_fig}
\end{figure}

Let us get back to the massive Yang--Mills theory (\ref{Euclid_action}).
We adopt the ansatz for $S^{1} \times \mathbb{R}^{3}$ space
\begin{equation}
g \mathscr{A}_{j}^{A} (\tau,\bm{x}) = \epsilon^{j A k} \frac{x^{k}}{r} \frac{1 - \widetilde{f} (r)}{r} , \ \ \ 
g \mathscr{A}_{4}^{A} (\tau,\bm{x}) = \frac{x^{A}}{r} \widetilde{a} (r) , \ \ \ 
\hat{\phi}^{A} (\tau,\bm{x}) = \frac{x^{A}}{r}
.
\label{E-ansatz}
\end{equation}
This ansatz has the same form as the ``static'' Julia--Zee ansatz on the ($3+1$)-dimensional Minkowski spacetime $\mathbb{R}^{1,3}$, which is obtained by the replacement $\widetilde{a} (r) \to - i \widetilde{a} (r)$ according to the Wick rotation. 
Therefore, we can use the arguments in the previous sections.

By introducing the dimensionless variable $\rho$ and functions $a$ and $f$ in the same way as (\ref{dimensionless}), the equations (\ref{YM_dyon_eq_a}) and (\ref{YM_dyon_eq_f}) are rewritten as
\begin{align}
& a^{\prime \prime} (\rho) + \frac{2}{\rho} a^{\prime} (\rho) - \frac{2}{\rho^{2}} a (\rho) f^{2} (\rho) =  0 , \label{E_eq_a2} \\
& f^{\prime \prime} (\rho) - \frac{1}{\rho^{2}} \left( f^{3} (\rho) - f (\rho) \right) - \left( a^{2} (\rho) + 1 \right) f (\rho) =  0 \label{E_eq_f2}
.
\end{align}
We set the boundary conditions of $a (\rho)$ and $f (\rho)$ as
\begin{align}
a (0) =  & 0 , \ \ \ a (\infty) = a_{\infty} , \label{boundary_a_E}\\
f (0) = & 1 , \ \ \ f (\infty) = 0 \label{boundary_f_E}
.
\end{align}
Notice that in the Euclidean space there is no restriction for the asymptotic value $a_{\infty}$ of $a (\rho)$, since the asymptotic form of $f (\rho)$ for large $\rho$
\begin{equation}
f (\rho) \approx F \exp \left\{ - \sqrt{1+ a_{\infty}^{2} } \rho \right\} 
,  \ \ \ (\rho \approx \infty)
\end{equation}
satisfies the boundary condition $f (\infty) = 0$ and exhibits no oscillating behavior for any values of $a_{\infty}$.
This differs from the dyons in the Minkowski spacetime.
We find that if $a (\rho)$ is a solution of the equations (\ref{E_eq_a2}) and (\ref{E_eq_f2}), then $- a(\rho)$ is also a solution of them.
Therefore, $a_{\infty}$ is restricted to take the nonnegative value $a_{\infty} \geq 0$ without loosing the generality.

Fig.~\ref{YM_dyon_Euclid_fig} is a plot of the solutions $a$ and $f$ of the Euclidean field equations (\ref{E_eq_a2}) and (\ref{E_eq_f2}) as functions of $\rho$ for various values of the asymptotic value $a_{\infty}$ of $a (\rho)$.
These solutions of the Yang--Mills dyon should be compared with the profile functions  (\ref{sd_sol_a}) and (\ref{sd_sol_f}) of the self-dual dyon.

The last term $- f(\rho)$ on the left hand side of (\ref{E_eq_f2}) originates from the kinetic term of the radially fixed scalar field $\hat{\bm{\phi}} (x)$, or equivalently, the gauge-invariant gluon mass term.
If the term $- f (\rho)$ is absent from the equation (\ref{E_eq_f2}), which means the absence of the gauge-invariant gluon mass term, the system (\ref{Euclid_action}) reproduces the pure massless Yang--Mills theory (\ref{S_E_YM}).
If $a_{\infty}$ is sufficiently large, $a_{\infty} \gg 1$, the term $- f (\rho)$ in (\ref{E_eq_f2}) can be negligible and hence the field equation (\ref{E_eq_f2}) can be approximated by (\ref{sd_eq_f}).
This means that the Yang--Mills dyon behaves as the self-dual dyon (\ref{sd_sol_a}) and (\ref{sd_sol_f}) for large $a_{\infty}$ except the neighborhood of the origin $\rho \approx 0$.
From (\ref{C_sd}), we observe that the ratio $C$ of the charges for the Yang--Mills dyon is equal to $1$ for large $a_{\infty}$:
\begin{equation}
C = \frac{q_{e}}{q_{m}}  \approx 1 , \ \ \ (a_{\infty} \gg 1).
\label{C_asymp_E}
\end{equation}

\begin{figure}[t]
\centering
\includegraphics[width=0.5\textwidth]{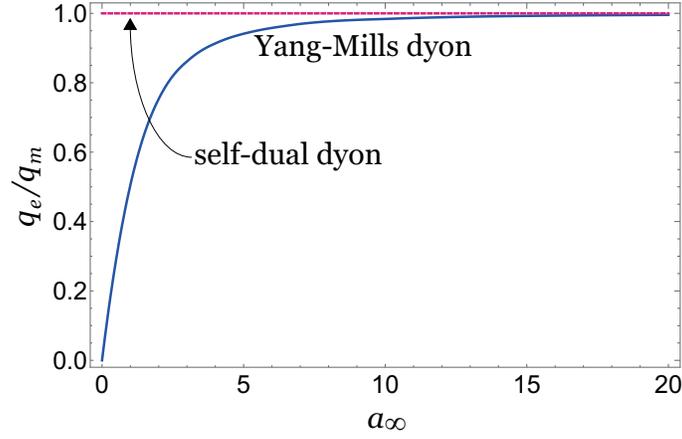}
\caption{The ratio of the charges $q_{e}/q_{m}$ as a function of $a_{\infty}$ for the fixed magnetic charge $q_{m} = 4 \pi / g$.
The broken line stands for the self-dual dyon, while the solid line stands for the Yang--Mills dyon.}
\label{YM_dyon_charge_Euclid}
\end{figure}

In the Yang--Mills dyon, the asymptotic value $a_{\infty}$ of the profile function $a (\rho)$ can be regarded as a function of the ratio $C$ of the charges $q_{e}$ and $q_{m}$, namely, $C = q_{e}/ q_{m}$.
Fig.~\ref{YM_dyon_charge_Euclid} is a plot of the ratio $C$ of the charges as a function of $a_{\infty}$ for the fixed magnetic charge $q_{m} = 4 \pi / g$.
We find that the electric charge $q_{e}$ of the Yang--Mills dyon depends on the asymptotic value $a_{\infty}$ of the profile function $a (\rho)$, while the electric charge of the self-dual dyon is fixed $q_{e} = q_{m}$.
We confirm numerically that the Yang--Mills dyon differs from the self-dual dyon for any finite $a_{\infty}$, but approaches the self-dual dyon in the limit $a_{\infty} \to \infty$.
This means that the upper bound of the electric charge $q_{e}$ exists and is given by
\begin{equation}
\biggl| \frac{q_{e}}{q_{m}} \biggr| \lesssim 1
.
\end{equation}

\begin{figure}[t]
\centering
\includegraphics[width=0.45\textwidth]{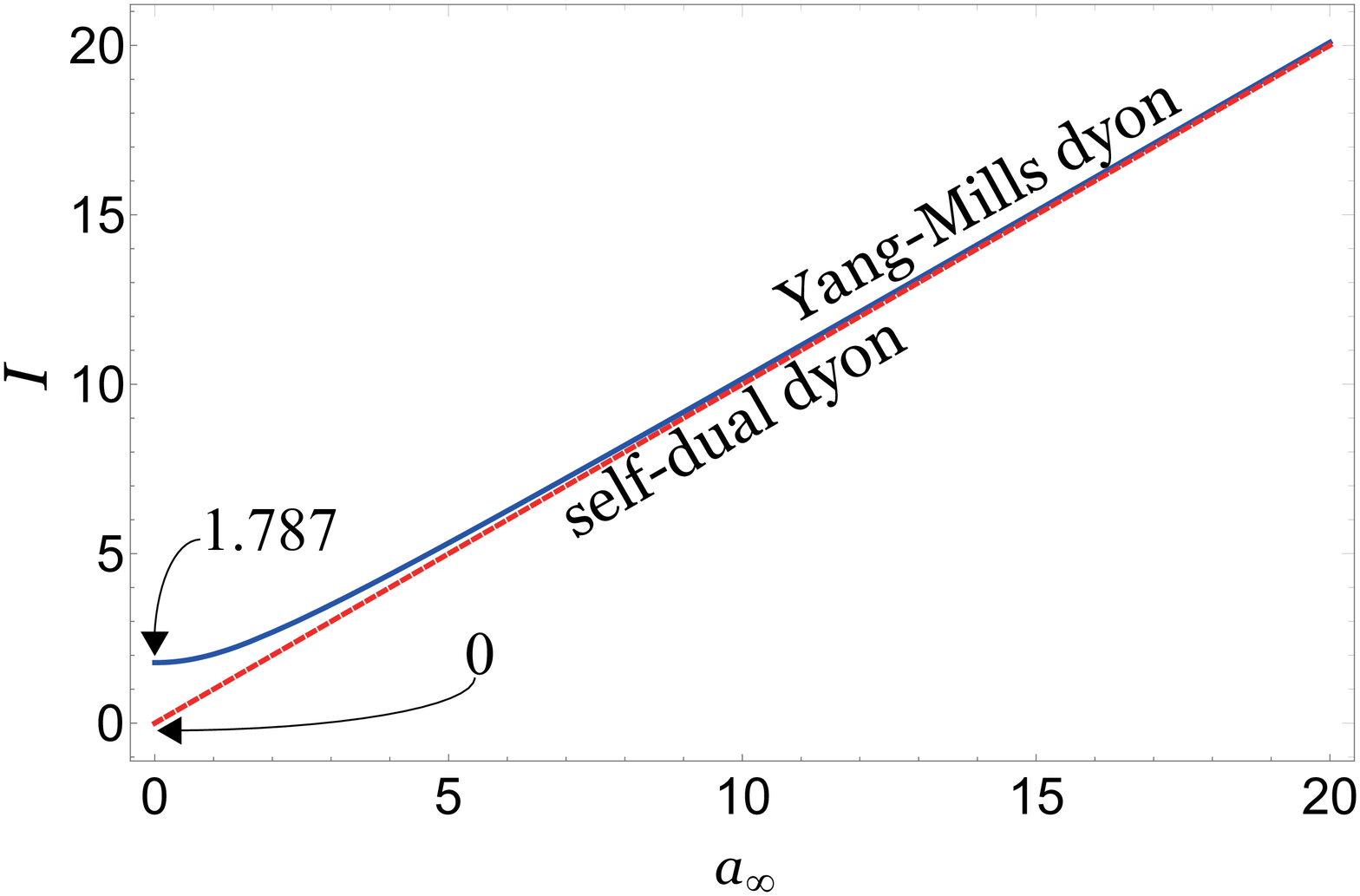} \ \ \ 
\includegraphics[width=0.45\textwidth]{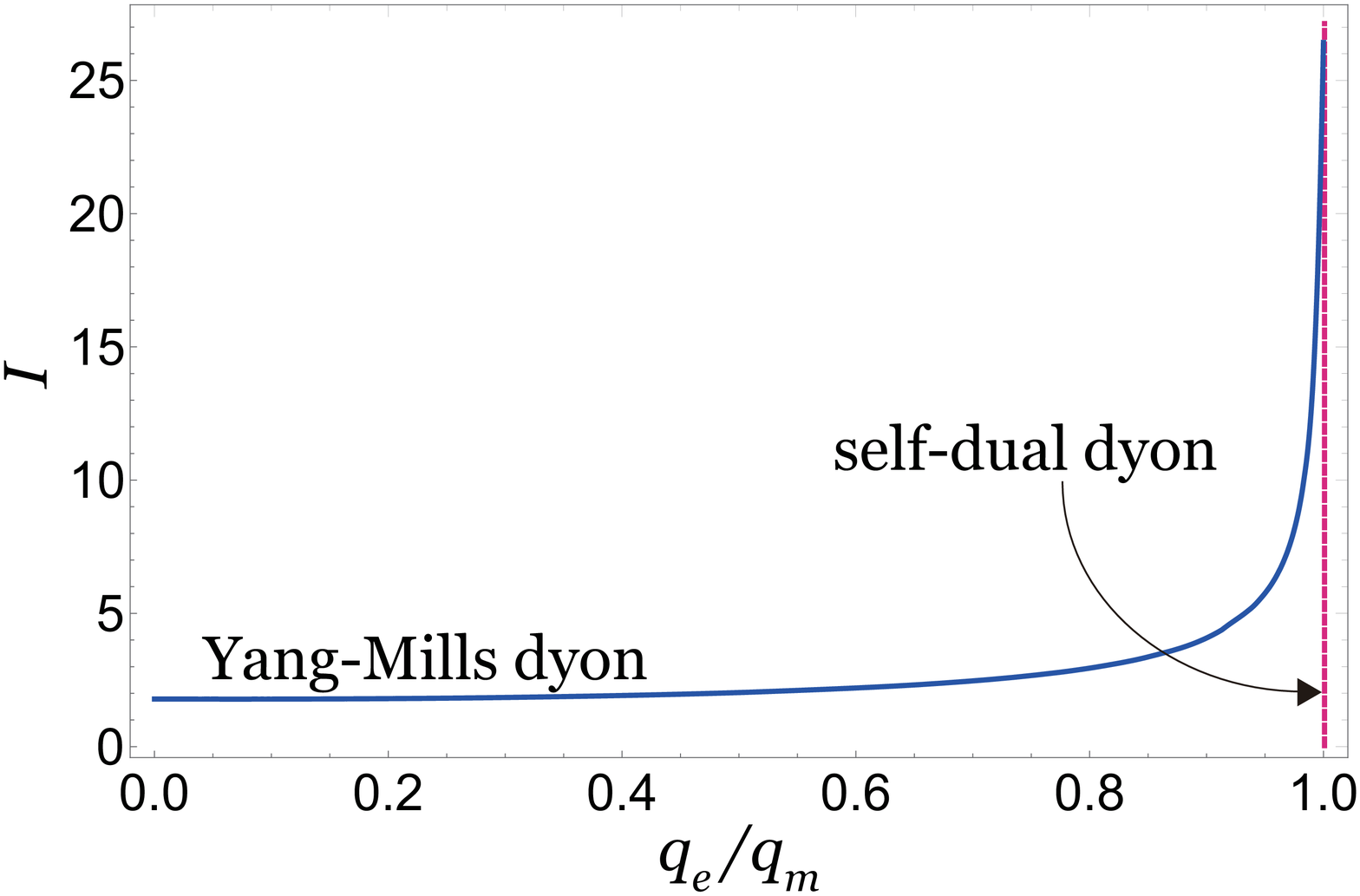}
\caption{The energy integral (\ref{energy_integral_E}) of the Yang--Mills dyon as a function of $a_{\infty}$ (left panel) and $q_{e}/q_{m}$ (right panel), which are denoted by the solid lines.
The broken lines are the energy integral (\ref{energy_integral_sd}) of the self-dual dyon as a  constituent of the KvBLL caloron. }
\label{YM_dyon_energy_Euclid_fig}
\end{figure}

The action (\ref{Euclid_action}) for a single Yang--Mills dyon is given by using (\ref{energy_integral}) and (\ref{energy_integral2})
\begin{equation}
S = \frac{4 \pi }{g^{2}} \frac{M_{\mathscr{X}}}{T} I \left( a_{\infty} \right) 
, \ \ \ M_{\mathscr{X}} := g v , 
\label{action-dyon}
\end{equation}
where we have defined the dimensionless energy integral $I (a_{\infty})$ by
\begin{align}
I (a_{\infty}) := & \int_{0}^{\infty} d \rho \ \biggl[ \frac{1}{2} \rho^{2} a^{\prime 2} (\rho) + a^{2} (\rho) f^{2} (\rho) + f^{\prime 2} (\rho) + \frac{( f^{2} (\rho) - 1)^{2}}{2 \rho^{2}}  + f^{2} (\rho)  \biggr]
,
\label{energy_integral_E}
\end{align}
which should be compared with (\ref{energy_integral_sd}) for the self-dual dyon.
The function $I (a_{\infty})$ of $a_{\infty}$ is monotonically increasing in $a_{\infty}$ with the lower bound $I (a_{\infty}) \geq I (a_{\infty} = 0) = 1.787$ based on the numerical calculations as given in the left panel of Fig.~\ref{YM_dyon_energy_Euclid_fig}.
The action for a KvBLL caloron is given by $S= 8 \pi^2/g^{2}$, which is $T$-independent.  This is not the case  for both the self-dual dyon (\ref{action_value_sd}) and the Yang--Mills dyon (\ref{action-dyon}). 

The $a_{\infty}$-dependence of $I (a_{\infty})$ is obtained as follows.
By taking the derivative of (\ref{energy_integral_E}) with respect to $a_{\infty}$, we have
\begin{align}
\frac{d I (a_{\infty})}{d a_{\infty}} = & \int_{0}^{\infty} d \rho \ \biggl[ 2 f^{\prime} (\rho) \frac{d f^{\prime} (\rho)}{d a_{\infty}} + 2 \left\{ \frac{1}{\rho^{2}} \left( f^{3} (\rho) - f (\rho) \right) + \left( a^{2} (\rho) + 1 \right) f (\rho) \right\} \frac{d f (\rho)}{d a_{\infty}} \nonumber\\
& \hspace{1.5cm}+ \rho^{2} a^{\prime} (\rho) \frac{d a^{\prime} (\rho)}{d a_{\infty}} + 2 a (\rho) f^{2} (\rho) \frac{d a (\rho)}{d a_{\infty}} \biggr] \nonumber\\
= & \biggl[ 2 f^{\prime} (\rho) \frac{d f (\rho)}{d a_{\infty}} \biggr] \biggl|_{\rho = 0}^{\rho = \infty} + \biggl[ \rho^{2} a^{\prime} (\rho) \frac{d a (\rho)}{d a_{\infty}} \biggr] \biggl|_{\rho = 0}^{\rho = \infty}  \nonumber\\
& + \int_{0}^{\infty} d \rho \ \biggl[ 2 \left\{ - f^{\prime \prime} (\rho) + \frac{1}{\rho^{2}} \left( f^{3} (\rho) - f (\rho) \right) + \left( a^{2} (\rho) + 1 \right) f (\rho) \right\} \frac{d f (\rho)}{d a_{\infty}} \nonumber\\
&\hspace{1.5cm}+ \left\{ - \rho^{2} a^{\prime \prime} (\rho) - 2 \rho a^{\prime} (\rho) + 2 a (\rho) f^{2} (\rho) \right\} \frac{d a (\rho)}{d a_{\infty}} \biggr] 
,
\label{I_a}
\end{align}
where we have integrated by parts.
The first term of (\ref{I_a}) vanishes, since the boundary conditions (\ref{boundary_f_E}) of $f (\rho)$ is $a_{\infty}$-independent.
The third term of (\ref{I_a}) also vanishes due to the field equations (\ref{E_eq_a2}) and (\ref{E_eq_f2}).
Thus, we obtain
\begin{equation}
\frac{d I (a_{\infty})}{d a_{\infty}} = \biggl[ \rho^{2} a^{\prime} (\rho) \frac{d a (\rho)}{d a_{\infty}} \biggr] \biggl|_{\rho = 0}^{\rho = \infty} = C (a_{\infty})
,
\label{I_diff_eq}
\end{equation}
where we have used the boundary conditions (\ref{boundary_a_E}) of $a (\rho)$.
By solving this equation with the initial condition $I (a_{\infty}  = 0) = I (0) = 1.787$, $I (a_{\infty})$ is written as
\begin{equation}
I (a_{\infty}) = I (0) + \int_{0}^{a_{\infty}} d s \ C (s)
.
\label{I_integral_eq}
\end{equation}
We find that from (\ref{C_asymp_E}) and (\ref{I_diff_eq}), $I (a_{\infty})$ behaves as
\begin{equation}
\frac{d I (a_{\infty})}{d a_{\infty}} \biggl|_{a_{\infty} \gg 1} = C ( a_{\infty} \gg 1) \approx 1
,
\end{equation}
and hence $I (a_{\infty})$ linearly diverges in $a_{\infty}$:
\begin{equation}
I (a_{\infty}) \approx a_{\infty} \to \infty , \ \ \ (a_{\infty} \gg 1).
\end{equation}
This is consistent with the numerical calculation shown in the left panel of Fig. \ref{YM_dyon_energy_Euclid_fig}.

Notice that the gauge field $\mathscr{A}_\mu$ of the Yang--Mills dyon has the asymptotic behavior dominated by $\mathscr{V}_\mu$, $\mathscr{A}_\mu \approx \mathscr{V}_\mu$,  at long distance, e.g., in the unitary gauge, 
\begin{align}
g \mathscr{A}_{r}^\prime \approx 0, \ \ 
\ g \mathscr{A}_{\theta}^\prime \approx 0 , \ \  
g \mathscr{A}_{\varphi}^\prime   \approx  \frac{1 - \cos \theta}{r \sin \theta} T_{3}   , \ \   
g \mathscr{A}_{4}^\prime \approx  \left( \widetilde{a}_{\infty} - \frac{C}{r} \right) T_{3}
, \ \   \left(r \gg (g v)^{-1} \right) ,
\label{asymptotic-YM}
\end{align}
which is the same as the asymptotic field  (\ref{asymptotic}) of a constituent dyon of the KvBLL caloron and hence the asymptotic field (\ref{asymptotic_sd}) of the self-dual dyon. 
Note that the self-dual dyon and Yang--Mills dyon no longer have $\tau$-dependence and hence they are trivially periodic, unlike the KvBLL caloron.

\subsection{The Yang--Mills dyon versus the self-dual dyon from the KvBLL caloron toward confinement/deconfinement phase transition}

In order to discuss the confinement/deconfinement phase transition in the Yang--Mills theory at finite temperature, we define the Polyakov loop operator $L (\bm{x})$ as
\begin{equation}
L (\bm{x}) := \frac{1}{\mathrm{tr} (\bm{1})} \mathrm{tr} \left\{ \mathscr{P} \exp \biggl[ i g \int_{0}^{T^{-1}} d \tau \ \mathscr{A}_{4} (\tau,\bm{x}) \biggr] \right\} 
,
\end{equation}
where $\mathscr{P}$ denotes the path-ordering prescription.
The asymptotic holonomy $\mathcal{P}_{\infty}$ is defined by the Polyakov loop operator at the spatial infinity
\begin{equation}
\mathcal{P}_{\infty} := \lim_{|\bm{x}| \to \infty} L (\bm{x})
.
\end{equation}
By performing the gauge transformation to the unitary gauge $\hat{\phi}^{A} (x) = \delta^{A 3}$, so that the ``time'' component $\mathscr{A}_{4} (x)$ of the gauge field becomes diagonal 
\begin{equation}
g \mathscr{A}_{4} (x) \equiv g \mathscr{A}_{4}^{A} (x) T_{A} =  \widetilde{a} (r) \frac{ \sigma_{3}}{2} =  gv {a} (\rho) \frac{ \sigma_{3}}{2}
,
\end{equation}
the asymptotic holonomy can be calculated as
\begin{align}
\mathcal{P}_{\infty} = & \lim_{|\bm{x}| \to \infty} \frac{1}{2} \mathrm{tr} \exp \biggl[ i \int_{0}^{T^{-1}} d \tau \ \widetilde{a} (r) \frac{\sigma_{3}}{2} \biggr] \nonumber\\
= & \lim_{r \to \infty} \frac{1}{2} \mathrm{tr} \exp \biggl[ \frac{i \widetilde{a} (r)}{2 T} \sigma_{3} \biggr]
= \lim_{r \to \infty} \cos \frac{\widetilde{a} (r)}{2 T} = \cos \biggl[ \frac{g v a_{\infty}}{2 T} \biggr]
,
\end{align}
where we have used (\ref{dimensionless}):
$
\widetilde{a} (\infty) = g v a (\infty) = g v a_{\infty}
$.
Fig. \ref{YM_dyon_charge_Euclid} implies that the asymptotic holonomy $\mathcal{P}_{\infty}$ depends on the electric charge $q_{e}$ through $a_{\infty}$:
\begin{equation}
\mathcal{P}_{\infty} (q_{e}) = \cos \biggl[ \frac{g v}{2T} a_{\infty} (q_{e}) \biggr]
,
\label{asymp_holonomy}
\end{equation}
since we have fixed the magnetic charge $q_{m}$ to the unit $q_{m} = 4\pi /g$.
In the limit of vanishing electric charge $q_{e} \to 0$, the Yang--Mills dyon reduces to the Yang--Mills monopole and the asymptotic holonomy $\mathcal{P}_{\infty}$ becomes trivial $\mathcal{P}_{\infty} \to 1$ according to $a_{\infty} \to 0$. 
In other words, the asymptotic holonomy $\mathcal{P}_{\infty}$ becomes nontrivial as long as the Yang--Mills dyon has a nonzero electric charge.

Note that this is not the case of the self-dual dyon. 
Since the electric charge $q_{e}$ of the self-dual dyon is fixed (\ref{C_sd}), the asymptotic holonomy $\mathcal{P}_{\infty}$ of the self-dual dyon does not depend on the electric charge
\begin{equation}
\mathcal{P}_{\infty} = \cos \biggl[ \frac{g v a_{\infty}}{2 T} \biggr] = \cos \frac{V}{2 T}
.
\end{equation}

The KvBLL calorons are extensively used to reproduce the confinement/deconfinement phase transition in the Yang--Mills theory at finite temperature \cite{Diakonov} by using the dyon gas model derived from the Yang--Mills theory where the interactions among dyons are calculated from quantum fluctuations around the dyon solution \cite{Diakonov2}. 
It should be remarked that the essential degrees of freedom responsible for confinement/deconfinement are not the KvBLL calorons themselves, but the constituent dyons characterized by the asymptotic behavior (\ref{asymptotic}) with nontrivial holonomy at spatial infinity. 

The (non-self-dual) Yang--Mills dyon obtained in this paper has also the nontrivial holonomy   and therefore can be used to explain the confinement/deconfinement phase transition in the Yang--Mills theory at finite temperature, instead of using the  KvBLL calorons or the self-dual dyons. 
In fact, it is possible to calculate the effective potential for the Polyakov loop in the same framework of the massive Yang--Mills model and show the existence of confinement/deconfinement transition at a certain critical temperature $T_c$, which is obtained as a definite ratio to the gluon mass $M$, see \cite{Kondo2015}. 
The evaluation assumes a non-vanishing uniform  background field for the component $\mathscr{A}_4$ and takes into account the quantum fluctuations to one-loop order. This procedure is regarded as the first approximation for the non-uniform gauge field $\mathscr{A}_4$ originating from the Yang--Mills dyon solution. 
Therefore, the existence of the Yang--Mills dyon with nontrivial holonomy (\ref{asymp_holonomy}) justify the the calculation of the effective potential of the Polyakov loop operator in a constant background \cite{Kondo2015} based on the same framework.

An advantage of using the Yang--Mills dyon is to give a successful explanation for quark confinement at zero temperature as the zero temperature limit of the finite temperature case. 
In the zero temperature limit $T \to 0$, the Yang--Mills dyon reduces to the Yang--Mills magnetic monopole \cite{Nishino}. It has been already shown that such Yang--Mills magnetic monopoles successfully explain quark confinement at zero temperature from the viewpoint of dual superconductivity, see e.g. \cite {KKSS} for a review. 
In contrast, the KvBLL caloron reduces to the BPST instanton in the zero temperature limit $T \to 0$. To the best of the author knowledge, however, the BPST instantons have not yet succeeded to explain quark confinement at zero temperature from the first principles without assuming additional inputs, see e.g.,  \cite{instanton-old-review}.

\section{Conclusion and discussion}

In this paper, we have constructed the dyon {\it configurations} in the pure $SU(2)$ Yang--Mills theory both in the $(3+1)$-dimensional Minkowski spacetime $\mathbb{R}^{1, 3}$ and in $S^{1} \times \mathbb{R}^{3}$ space by incorporating a gauge-invariant gluon mass term  even in the absence of the scalar field. 
Such a gauge-invariant mass term is obtained through a gauge-independent description of the BEH mechanism proposed in \cite{Kondo2016}. 
The procedure for obtaining the relevant dyon is guided by the ``complementarity'' between the $SU(2)$ gauge-adjoint scalar model with a single radially fixed scalar field and the massive $SU(2)$ Yang--Mills theory.
In fact, we have obtained the static and spherically symmetric dyon configuration in the $SU(2)$ massive Yang--Mills theory by solving the field equations of the ``complementary'' $SU(2)$ gauge-adjoint scalar model with a single radially fixed scalar field.
We have found that the static energy or the rest mass of the obtained Yang--Mills dyon is finite and proportional to the mass $M_{\mathscr{X}}$ of the Yang--Mills gauge field $\mathscr{A}$ representing the existence of the massive component  $\mathscr{X}$.

In the long-distance region, we observed that the Yang--Mills dyon configuration $\mathscr{A}$ reduce to the restricted field $\mathscr{V}$, which agrees with the dyonic extension of the Wu--Yang magnetic monopole as a consequence of the suppression of the massive modes $\mathscr{X}$ in the long-distance region.
This feature is similar to the usual Julia--Zee dyons.
In the short-distance region, on the other hand, the Wu--Yang magnetic monopole becomes singular, while  the Julia--Zee dyon remains non-singular even at the origin. 
In the Yang--Mills dyon, we found that the massive components $\mathscr{X}$ play the very important role of canceling the singularity of $\mathscr{V}$ in the short-distance region such that the original gauge field $\mathscr{A}$ remains non-singular at the origin.
This regularity of the Yang--Mills dyon is guaranteed by the logarithmic behavior of the gauge field itself without the aid of the scalar field, which vanishes at the origin as seen in Julia--Zee dyons. 
This behavior renders the energy of the Yang--Mills dyon finite even if the magnitude of the scalar field is fixed. 
It should be remarked that the chromomagnetic field $\mathscr{B}$ is divergent at the origin due to the logarithmic behavior of the solution $f (\rho)$, which is, however, unessential for obtaining finite physical quantities such as energy, magnetic and electric charge density, and magnetic flux.
Moreover, in the Yang--Mills dyon configuration, the time-component of the gauge field $\mathscr{A}_{0} $ is regular, whose regularity is supported by  the absence of the time-component of the high-energy massive mode: $\mathscr{X}_{0} \equiv 0$.


Furthermore, we estimated the static mass of the Yang--Mills dyon by using the values of the previous studies \cite{Shibata2007, Bloch2003}.
We found that the heaviest static mass of the Yang--Mills dyon $M_{\rm dyon} \approx 1.18 \mathrm{GeV}$ is around the off-diagonal gluon mass $M_{\mathscr{X}} \approx 1.2 \mathrm{GeV}$.
This is a quite reasonable result for quark confinement to be realized due to condensation of the relevant Yang--Mills monopoles according to the dual superconductor picture.
We need, however, more careful investigations to conclude whether or not the interactions among monopoles are indeed sufficient for realizing the monopole condensations, as examined by Polyakov \cite{Polyakov} in the three dimensional case.

We observed that the Yang--Mills dyon cannot acquire the electric charge which is equal to the magnetic one.
This is caused by a gauge-invariant mass term.
In the contexts of instantons, the electric charge is equal to the magnetic one by definition, i.e., the (anti-)self-dual condition. However, there do not exist such (anti-)self-dual objects in our theory due to the mass term.
We found that the Yang--Mills dyon in $S^{1} \times \mathbb{R}^{3}$ space has a nontrivial holonomy.
This implies that our (non-self-dual) dyon with nontrivial holonomy $\mathcal{P}_{\infty}$ can be used to explain the confinement/deconfinement phase transition in the Yang--Mills theory at finite temperature based on the dual superconductor picture for confinement, instead of using the traditional KvBLL calorons or the self-dual dyon.

\begin{table}[t]
\centering
\begin{tabular}{|c||c|c|}
\hline
 & KvBLL caloron & ``Yang--Mills caloron'' \\ \hline\hline
self-dual & Yes & No \\ \hline
constituents & (anti-)self-dual dyons & Yang--Mills dyons \\ \hline
$T \to 0$ & BPST instanton & Yang--Mills monopole-antimonopole chain \\ \hline
$V, v \to 0$ & HS caloron & Shnir's caloron \\ \hline
\end{tabular}
\caption{Properties of the KvBLL caloron and the Yang--Mills caloron for comparison.
The Shnir's caloron found in \cite{Shnir2007} contains the HS caloron in a minimal topological charge sector. 
Moreover, in the limit $V \to 0$, the massive Yang--Mills theory  reduces to the ordinary  massless Yang--Mills theory, since the scalar field $\phi$ decouples.}
\label{Table}
\end{table}

Finally, we give a conjecture that there will exist a caloron in the massive Yang--Mills theory so that the (non-self-dual) Yang--Mills dyon found in this paper could be identified with a constituent of the  caloron. We call such a caloron the {\it Yang--Mills caloron}.
In \cite{Shnir2007}, the non-self-dual calorons which have the axial symmetry were constructed in the pure massless $SU(2)$ Yang--Mills theory in the four-dimensional Euclidean space.
Such axially symmetric solutions were also found in the Yang--Mills--Higgs model, i.e., the radially variable model by adopting the Kleihaus--Kunz ansatz \cite{Kleihaus-Kunz-Shnir}.
We are therefore led to consider the axially symmetric Yang--Mills calorons in the massive Yang--Mills theory.
In the zero temperature limit $T \to 0$, the Yang--Mills dyon reduces to the Yang--Mills magnetic monopole \cite{Nishino}.
This property is expected to hold in the Yang--Mills calorons, which will reduce to the Yang--Mills monopole-antimonopole chains.
See Table \ref{Table} for the properties of the KvBLL caloron and the conjectured Yang--Mills caloron.
This issue will be explored in near future.

\section*{Acknowledgement}
This work was supported by Grant-in-Aid for Scientific Research, JSPS KAKENHI Grant Number (C) No.19K03840.


\begin{thebibliography}{99}

\bibitem{dual-superconductor}
Y. Nambu, Phys. Rev. D{\bf 10}, 4262 (1974). 
\\
G. 't Hooft, in: High Energy Physics, edited by A. Zichichi (Editorice Compositori, Bologna,
1975).
\\
S. Mandelstam, Phys. Rep. {\bf 23}, 245 (1976).


\bibitem{tHP}
G. 't Hooft, Nucl. Phy. B{\bf 79}, 276 (1974). 
\\
A.M. Polyakov, Sov. Phys. - JETP{\bf 41}, 988 (1975),
JETP Lett.{\bf 20}, 194 (1974).

\bibitem{KKSS}
K.-I. Kondo, S. Kato, A. Shibata and T. Shinohara, 
Phys. Rept. {\bf 579}, 1 (2015).
arXiv:1409.1599 [hep-th]


\bibitem{Nishino}
S. Nishino, R. Matsudo, M. Warschinke, and K.-I. Kondo, PTEP{\bf 2018}, 103B04 (2018).

\bibitem{Kondo2016}
K.-I. Kondo, Phys. Lett. B{\bf 762}, 219 (2016).
arXiv:1606.06194 [hep-th]

\bibitem{Kondo2018}
K.-I. Kondo, 
Eur. Phys. J. C78,  577 (2018).
arXiv:1804.03279 [hep-th] 


\bibitem{BEH}
P.W. Higgs,
Phys. Lett. \textbf{12},  132
 (1964).
P.W. Higgs,
Phys. Rev. Lett.  \textbf{13},  508
 (1964).
\\
F. Englert and R. Brout,
Phys. Rev. Lett. \textbf{13}, 321
 (1964). 
\\
G.S. Guralnik, C.R. Hagen, and T.W.B. Kibble,
Phys. Rev. Lett. \textbf{13}, 585 (1964). 


\bibitem{Kondo-Murakami-Shinohara}
K.-I. Kondo, T. Murakami, and T. Shinohara, Eur. Phys. J. C{\bf 42}, 475 (2005).

\bibitem{Julia--Zee}
B. Julia and A. Zee, Phys. Rev. D{\bf 11}, 2227 (1975).



\bibitem{textbooks}
N. Manton and P. Sutcliffe,
{\it Topological Solitons},
(Cambridge University Press, Cambridge, 2004).
\\
E.J. Weinberg,
{\it Classical Solutions in Quantum Field Theory},
(Cambridge University Press, Cambridge, 2012).
\\
Ya. Shnir,
{\it Magnetic Monopoles},
(Springer, Berlin, 2005).



\bibitem{Wu--Yang}
T.T. Wu and C.N. Yang, Phys. Rev. D{\bf 12}, 3845 (1975).

\bibitem{KvBLL}
T.C. Kraan and P. van Baal, Phys. Lett. B{\bf 435}, 389 (1998).
\\
T.C. Kraan and P. van Baal, Nucl. Phys. B{\bf 533}, 627 (1998).
\\
K. Lee and C. Lu, Phys. Rev. D{\bf 58}, 025011 (1998).

\bibitem{Diakonov}
D. Diakonov, Nucl. Phys. B (Proc. Suppl.) {\bf 195}, 5 (2009).
\\
M.A. Lopez-Ruiz, Y. Jiang, and J. Liao, Phys. Rev. D{\bf 97}, 054026 (2018).



\bibitem{Bais-Primack}
F. A. Bais and J. R. Primack,
Phys. Rev. D{\bf 13}, 819 (1976).

\bibitem{Shibata2007}
A. Shibata, S. Kato, K.-I. Kondo, T. Murakami, T. Shinohara, and S. Ito, Phys. Lett. B{\bf 653}, 101 (2007).
arXiv:0706.2529 [hep-lat]

\bibitem{Bloch2003}
J.C.R. Bloch, Few Body Syst.{\bf 33}, 111 (2003).
\\
J.C.R. Bloch, A. Cucchieri, K. Langfeld, and T. Mendes
Nucl. Phys. B{\bf 687}, 76--100 (2004).  
e-Print: hep-lat/0312036 



\bibitem{BPST}
A. Belavin, A. Polyakov, A. Schwartz, and Yu. Tyupkin, Phys. Lett. {\bf 59B}, 85 (1975).


\bibitem{Harrington-Shepard}
B.J. Harrington and H.K. Shepard, Phys. Rev. D{\bf 17}, 2122 (1978); {\bf 18}, 2990 (1978).

\bibitem{Diakonov2}
D. Diakonov, N. Gromov, V. Petrov, and S. Slizovskiy, Phys. Rev. D{\bf 70}, 036003 (2004).


\bibitem{Kondo2015}
K.-I. Kondo, arXiv:1508.02656 [hep-th].


\bibitem{instanton-old-review}
T. Schafer and E.V. Shuryak,
Rev. Mod. Phys. {\bf 70}, 323--426 (1998).  
e-Print: hep-ph/9610451   


\bibitem{Polyakov}
A.M. Polyakov,
Nucl. Phys. B{\bf 120}, 429  (1977).


\bibitem{Shnir2007}
Ya. Shnir, Europhys. Lett. {\bf 77}, 21001 (2007).


\bibitem{Kleihaus-Kunz-Shnir}
B. Hartmann, B. Kleihaus, and J. Kunz, Mod. Phys. Lett. A{\bf 15}, 1003 (2000). \\
B. Kleihaus and J. Kunz, Phys. Rev. D{\bf 61}, 025003 (1999).
\\
B. Kleihaus, J. Kunz, and Ya. Shnir, Phys. Lett. B{\bf 570}, 237 (2003);
Phys. Rev. D{\bf 68}, 101701 (2003);
Phys. Rev. D{\bf 70}, 065010 (2004).



\end{thebibliography}
\end{document}